\documentclass[apj]{emulateapj}

\newcommand{\lsim}{\raisebox{-.4ex}{$\stackrel{<}{\scriptstyle \sim}$}}

\shorttitle{
Diffuse X-ray Emission from Planetary Nebulae with Nebular O~{\sc vi}}  
\shortauthors{Ruiz et al.}

\begin{document}


\title{
Detection of Diffuse X-ray Emission from Planetary Nebulae with 
Nebular O~{\sc vi}}


\author{
N.\ Ruiz\altaffilmark{1}, 
Y.-H.\ Chu\altaffilmark{2}, 
R.A.\ Gruendl\altaffilmark{2}, 
M.A.\ Guerrero\altaffilmark{1}, 
R.\ Jacob\altaffilmark{3}, 
D.\ Sch\"onberner\altaffilmark{3}, 
\& 
M.\ Steffen\altaffilmark{3}
} 

\email{nieves@iaa.es}


\altaffiltext{1}{
Instituto de Astrof\'\i sica de Andaluc\'\i a, IAA-CSIC, 
c/ Glorieta de la Astronom\'\i a s/n, 18008 Granada, Spain}
\altaffiltext{2}{
Department of Astronomy, University of Illinois, 
1002 West Green Street, Urbana, IL 61801, USA} 
\altaffiltext{3}{
Leibniz-Institut f\"ur Astrophysik Potsdam (AIP), 
An der Sternwarte 16, 14482 Potsdam, Germany
}


\begin{abstract}
The presence of O~{\sc vi} ions can be indicative of plasma temperatures
of a few $\times$10$^5$ K that is expected in heat conduction layers 
between the hot shocked stellar wind gas at several 10$^6$~K and the 
cooler (10$^4$~K) nebular gas of planetary nebulae (PNe).
We have used \emph{FUSE} observations of PNe to search for nebular 
O~{\sc vi} emission or absorption as a diagnostic of conduction 
layer to ensure the presence of hot interior gas.  
Three PNe showing nebular O~{\sc vi}, namely IC\,418, NGC\,2392, and 
NGC\,6826, have been selected for \emph{Chandra} observations and 
diffuse X-ray emission is indeed detected in each of these PNe.  
Among the three, NGC\,2392 has peculiarly high diffuse X-ray luminosity and 
plasma temperature compared with those expected from its stellar wind's
mechanical luminosity and terminal velocity. 
The limited effects of heat conduction on the plasma temperature 
of a hot bubble at the low terminal velocity of the stellar wind 
of NGC\,2392 may partially account for its high plasma temperature, 
but the high X-ray luminosity needs to be powered by processes 
other than the observed stellar wind, probably caused by the 
presence of an unseen binary companion of the CSPN of NGC\,2392.  
We have compiled relevant information on the X-ray, stellar, and nebular 
properties of PNe with a bubble morphology and found that the expectations 
of bubble models including heat conduction compare favorably with the 
present X-ray observations of hot bubbles around H-rich CSPNe, but have 
notable discrepancies 
for those around H-poor [WR] CSPNe.  
We note that PNe with more massive central stars can produce hotter 
plasma and higher X-ray surface brightness inside central hot bubbles.
 
\end{abstract}



\section{Introduction}

Planetary nebulae (PNe) consist of the stellar material ejected by 
low- and intermediate-mass stars near the end of their evolution, 
before turning into white dwarfs.  
The physical structure of a PN is largely determined by the photoionization 
of the slow, dense wind ejected during the asymptotic giant branch (AGB) 
phase by the intense stellar radiation field and its interaction with the 
subsequent fast, tenuous wind emanating from the hot stellar core at  
terminal velocities up to $\sim$4,000 km~s$^{-1}$ \citep{CRP85,GRM10}.  
In this interacting stellar-wind model \citep{Kwok83}, the physical 
structure of a PN is similar to that of a wind-blown bubble and 
will comprise (1) a central cavity filled with shocked fast wind 
at temperatures of 10$^7$--10$^8$ K, (2) a dense shell of swept-up 
AGB wind at 10$^4$ K (the bright ring seen in optical images), and 
(3) an outer envelope of ionized AGB wind material reshaped by the 
leading shock set up by ionization 
\citep[cf.][]{S-VK87,MS91,Mellema94,VMG-S02,Perinotto_etal04}.
At the interface between the shocked fast wind and the swept-up AGB 
wind, a contact discontinuity forms and heat conduction is expected 
to occur \citep{Spitzer62}.  
The resulting mass evaporation from the dense nebular shell into 
the hot interior lowers the temperature and raises the density 
of the hot gas \citep{WR77}, significantly increasing the X-ray 
emissivity. 
Hydrodynamic models of PNe with heat conduction predict diffuse 
X-ray emission that should be easily detectable with modern X-ray 
observatories \citep[e.g.,][]{VK85,ZP98,Schon_etal06,Steffen_etal08}.

\emph{Chandra} and \emph{XMM-Newton} observations of diffuse X-ray emission 
from PNe have been used to investigate the physical properties of the hot 
interior gas, which shows plasma temperatures of (1--3)$\times$10$^6$ K and 
X-ray luminosities $L_{\rm X}$=2$\times$10$^{30}$--3$\times$10$^{32}$ 
erg~s$^{-1}$ 
\citep[e.g.,][]{CHU04,Guerrero_etal2005,MTZ05,KN08,Ruiz_etal11}. 
The X-ray morphology, low plasma temperatures, and moderate $L_{\rm X}$ 
are quantitatively consistent with those expected from bubble models 
with heat conduction \citep{Steffen_etal08}.

\begin{deluxetable*}{lclllrc}
\tabletypesize{\scriptsize}
\tablewidth{0pt}
\tablecaption{
Space Observations of IC\,418, NGC\,2392, and NGC\,6826
\vspace*{-0.2cm}
}
\tablehead{
\multicolumn{7}{c}{\emph{FUSE} UV Observations}
}
\startdata
\multicolumn{1}{c}{Object} & 
\multicolumn{1}{c}{Program ID} & 
\multicolumn{1}{c}{Aperture} & 
\multicolumn{1}{c}{Date} & 
\multicolumn{1}{c}{Processing Version} & 
\multicolumn{1}{c}{} & 
\multicolumn{1}{c}{$t_{\rm exp}$} \\ 
\multicolumn{1}{c}{} & 
\multicolumn{1}{c}{} & 
\multicolumn{1}{c}{} & 
\multicolumn{1}{c}{} & 
\multicolumn{1}{c}{} & 
\multicolumn{1}{c}{} & 
\multicolumn{1}{c}{[s]} \\
\hline
IC\,418   & LWRS & ~~~~~~~~~P1151111 & 2001 Dec 2  & ~CalFUSE v3.2.3 & & 4,440 \\
NGC\,2392 & LWRS & ~~~~~~~~~B0320601 & 2001 Feb 21 & ~CalFUSE v3.2.3 & & 2,620 \\
NGC\,6826 & MDRS & ~~~~~~~~~P1930401 & 2000 Aug 8  & ~CalFUSE v3.2.3 & & 5,800 \\
\hline
\hline
\vspace*{-0.2cm} \\
\multicolumn{7}{c}{\emph{Chandra} X-ray Observations} \\
\vspace*{-0.2cm} \\
\hline
\vspace*{-0.2cm} \\
\multicolumn{1}{c}{Object} & 
\multicolumn{1}{c}{Observation ID} & 
\multicolumn{1}{c}{Instrument \& Pointing} & 
\multicolumn{1}{c}{Date} & 
\multicolumn{1}{c}{Processing Version} & 
\colhead{$t_{\rm obs}$}  & 
\colhead{$t_{\rm exp}$}  \\
\multicolumn{1}{c}{} & 
\multicolumn{1}{c}{} & 
\multicolumn{1}{c}{} & 
\multicolumn{1}{c}{} & 
\multicolumn{1}{c}{} & 
\multicolumn{1}{c}{[ks]} & 
\multicolumn{1}{c}{[ks]} \\
\hline
IC\,418    & 7440 & ~~~~~~~~ACIS-S S3 & 2006 Dec 12 & ~~~~~DS~7.6.9  & 51.1 & 50.4 \\
NGC\,2392  & 7421 & ~~~~~~~~ACIS-S S3 & 2007 Nov 13 & ~~~~~DS~7.6.11 & 58.1 & 57.4 \\
NGC\,6826  & 7439 & ~~~~~~~~ACIS-S S3 & 2007 Jun 11 & ~~~~~DS~7.6.10 & 34.5 & 34.1 \\
           & 8559 & ~~~~~~~~ACIS-S S3 & 2007 Jul 24 & ~~~~~DS~7.6.10 & 15.0 & 14.9 \\
\hline
\hline
\vspace*{-0.2cm} \\
\multicolumn{7}{c}{\emph{HST} Optical Observations} \\
\vspace*{-0.2cm} \\
\hline
\vspace*{-0.2cm} \\
\multicolumn{1}{c}{Object} & 
\multicolumn{1}{c}{Program ID} & 
\multicolumn{1}{c}{Instrument \& Pointing} & 
\multicolumn{1}{c}{Date} & 
\multicolumn{1}{c}{Filter} & 
\multicolumn{1}{c}{} & 
\multicolumn{1}{c}{$t_{\rm exp}$} \\ 
\multicolumn{1}{c}{} & 
\multicolumn{1}{c}{} & 
\multicolumn{1}{c}{} & 
\multicolumn{1}{c}{} & 
\multicolumn{1}{c}{} & 
\multicolumn{1}{c}{} & 
\multicolumn{1}{c}{[s]} \\
\hline
IC\,418   & 8773 & ~~~~~~~WFPC2-PC  & 2001 Oct  3 & ~~~~~~~~F502N & & 600 \\
NGC\,2392 & 8499 & ~~~~~~~WFPC2-WF3 & 2000 Jan 11 & ~~~~~~~~F656N & & 100 \\
NGC\,6826 & 6117 & ~~~~~~~WFPC2-PC  & 1996 Jan 27 & ~~~~~~~~F502N & & 100
\enddata 
\label{obs}
\end{deluxetable*}

To make rigorous comparisons between observations and model predictions
of hot gas in PN interiors and to achieve statistical significance in the 
comparisons, the small sample of PNe with known diffuse X-ray emission 
needs to be enlarged.
While it has been observed that PNe possessing detectable diffuse X-ray 
emission exhibit sharp inner shells of swept-up AGB wind, the reverse may
not be true \citep{Kastner_etal12}.  
A more reliable diagnostic to ensure the detection of diffuse X-ray
emission from PNe is needed.
To this end, we have utilized the collisionally ionized O~{\sc vi} 
expected in interface layers to diagnose the presence of hot gas.
Using archival \emph{Far Ultraviolet Spectroscopic Explorer} 
\citep[\emph{FUSE};][]{Metal00} observations of PNe, we find three 
cases, IC\,418, NGC\,2392, and NGC\,6826,
exhibiting nebular O~{\sc vi} absorption lines.  
We have subsequently obtained \emph{Chandra} observations of these
three PNe, and indeed diffuse X-ray emission is detected in all
three cases.
In this paper, we describe the \emph{FUSE} observations of PNe in 
Section 2, report the \emph{Chandra} observations of the three PNe 
in Section 3, discuss the X-ray results in the framework of models 
of PNe formation and evolution in Section 4, and summarize the main 
conclusions in Section 5.

\section{O~{\sc vi} Diagnostic of Hot Gas}

A PN with hot interior gas will possess an interface layer in
which the temperature falls from 10$^6$ K near the hot gas to 
10$^4$ K in the cool nebular shell.  
Thermal collisions in the interface layer produce highly ionized species, 
such as C~{\sc iv}, N~{\sc v}, and O~{\sc vi}, whose excitation potentials 
are 47.8, 77.5, and 113.9 eV, and whose fractional abundances peak at 
$\sim$1$\times$10$^5$, 2$\times$10$^5$, and 3$\times$10$^5$ K, respectively 
\citep{SvS82}.
Cloudy simulations \citep{Ferland_etal98} show, however, that such 
ions can also be produced by photoionization in appreciable amounts 
for luminous central stars of planetary nebulae (CSPNe) hotter than 
$\sim$35,000, 85,000, and 140,000 K, respectively.  
Being the hardest to be produced by photoionization, O~{\sc vi} is thus 
the best choice among the three species for diagnosing interface 
layers and hot gas in PNe.
O~{\sc vi} has a resonance doublet at $\lambda\lambda$1031.9,1037.6 
\AA, a primary spectral feature targeted by \emph{FUSE}.

\begin{figure*}[!ht]
\centering
\includegraphics[width=1.70\columnwidth]{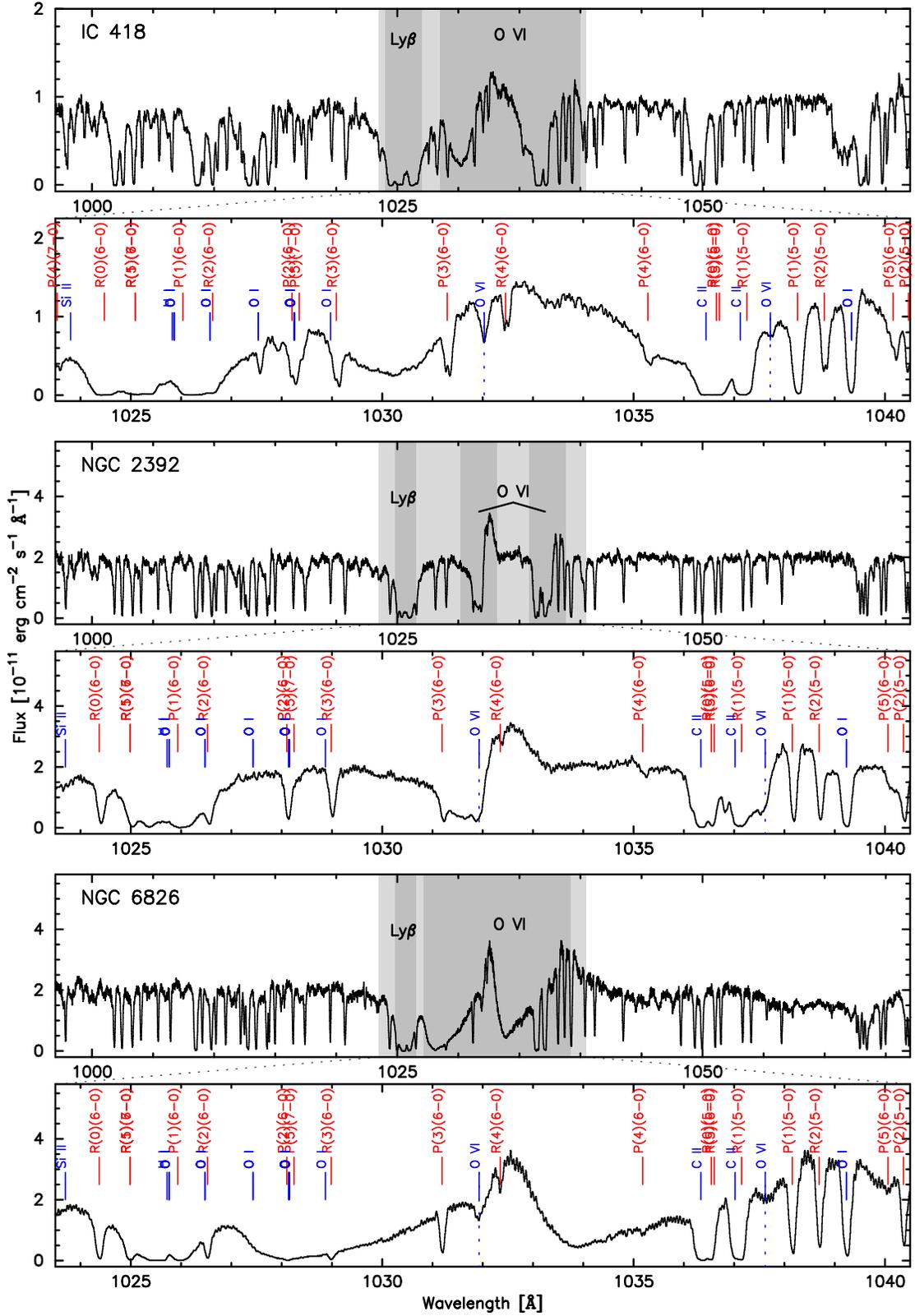}
\caption[]{
\emph{FUSE} spectra showing the region around the O~{\sc vi} lines
for the central stars of IC\,418 (top panels), NGC\,2392
(middle panels), and NGC\,6826 (bottom panels).  
For each object the upper panel shows a wide spectral range with 
the broad Ly$\beta$ and O~{\sc vi} P~Cygni profiles highlighted 
(darker grey).  
The shaded region in the upper panel is expanded to better show narrow
absorption features superposed on the broader spectrum.  Absorption
from the H$_2$ Lyman and Werner bands are marked in red and ionic
transition are marked with blue.  The narrow, nebular O~{\sc vi}
components are also marked with vertical dashed lines.}
\label{FUSE_fig}
\end{figure*}

The O~{\sc vi} doublet from the interface layer can be detected 
as emission features in \emph{FUSE} spectra if the entrance 
aperture does not include the CSPN, or as absorption features 
if the entrance aperture includes the CSPN.
\emph{FUSE} observations of NGC\,6543 and NGC\,7009, two PNe 
with diffuse X-ray emission, indeed detected nebular O~{\sc vi}
emission from the interface layers \citep{GRU04,ISC02}.

\begin{figure*}
\epsscale{1.0}
\includegraphics[angle=0,scale=1.0]{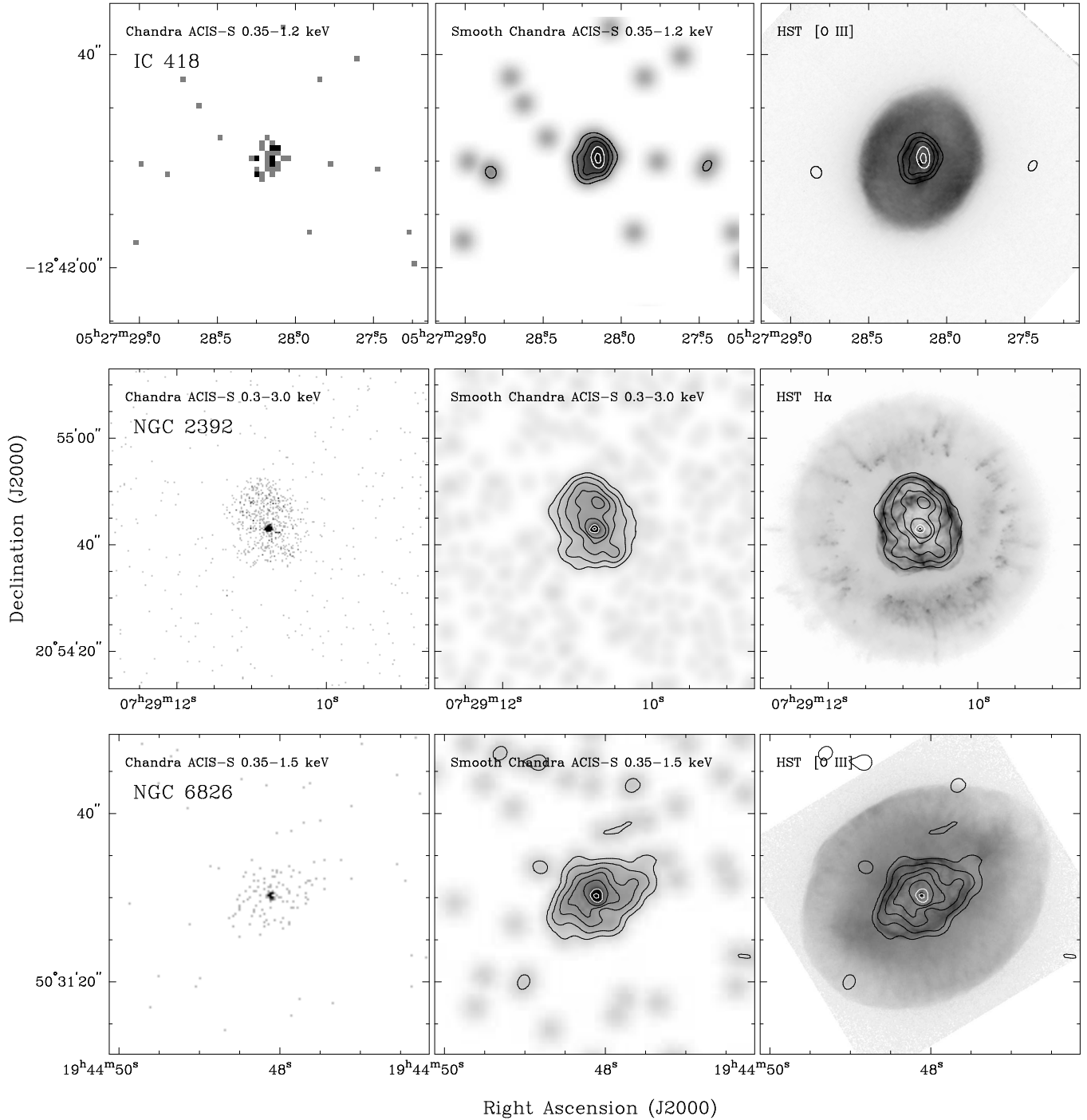}
\caption{
\emph{Chandra} ACIS-S raw ({\it left}) and smoothed ({\it center}) X-ray 
images and \emph{HST} optical narrow-band images ({\it right}) of IC\,418 
({\it top row}), NGC\,2392 ({\it middle row}), and NGC\,6826 ({\it bottom 
row}).  
The X-ray contours extracted from the smoothed X-ray images have 
been overlaid onto the optical images.  
The black and grey (only for IC\,418) contours mark the X-ray levels 
corresponding to 10\%, 25\%, 50\%, 75\%, and 95\% of the peak intensity 
of the diffuse emission.  
Two additional white contours defining the 25\% and 75\% of the peak 
intensity of the central source are shown for NGC\,2392 and NGC\,6826.  
}
\label{img}
\end{figure*}

The \emph{FUSE} archive contains high-dispersion spectra 
\citep[$\lambda/\Delta\lambda \approx $20,000, ][]{Sahnow_etal00} 
of a large number of CSPNe.
These observations provide two pieces of information pertinent to the 
hot interior gas.  
First, the stellar O~{\sc vi} lines may show P~Cygni profiles 
indicating the existence of a fast stellar wind that is needed
to produce the hot interior gas in a PN.
Second, there may be narrow nebular O~{\sc vi} absorption features
superposed on the stellar P~Cygni profile, and if the CSPN is 
cooler than 140,000 K, the nebular absorption must originate 
from an interface layer, requiring the existence of hot interior gas.  
\emph{FUSE} makes simultaneous observations through the 
30\arcsec$\times$30\arcsec\ low (LWRS), 
4\arcsec$\times$20\arcsec\ medium (MDRS), and 
1\farcs25$\times$20\arcsec\ high (HIRS) 
resolution apertures, that are offset from one 
another.  
It is possible that some \emph{FUSE} observations of CSPNe had one
or two off-source apertures falling within the nebula, allowing us
to search for O~{\sc vi} emission from the interface layer.

We have used \emph{FUSE} observations of the nebular shells of a 
dozen PNe to search for nebular O~{\sc vi} emission, and analyzed 
\emph{FUSE} observations of $\simeq$60 CSPNe to search for narrow 
nebular O~{\sc vi} absorption lines in their spectra. 
Each observation is typically composed of multiple spectra.  
The individual exposures were processed using the CalFUSE v3.2.3 
\citep{Dixon_etal07} to obtain an extracted spectrum.  
Prior to combining the individual spectra, the relative alignment 
of each was checked by performing a cross correlation. 
Typical offsets were less than a few km~s$^{-1}$ for the LiF1A 
module. 
The spectra were then combined using a weighted average 
after accounting for the small offsets in wavelength.

Among the PNe with \emph{FUSE} observations, IC\,418, NGC\,2392, 
and NGC\,6826 display narrow nebular O~{\sc vi} absorption lines 
superposed on the spectra of their CSPNe (Figure~\ref{FUSE_fig}, 
see Table~\ref{obs} for the details of these observations).  
We have thus obtained \emph{Chandra} X-ray observations of these three PNe.
Note that the \emph{Chandra} observation of NGC\,2392 was a follow-up
of an earlier \emph{XMM-Newton} observation \citep{Guerrero_etal2005}.

\begin{figure}[!t]
\centering
\epsscale{1.0}
\includegraphics[angle=0.0,width=0.76\columnwidth]{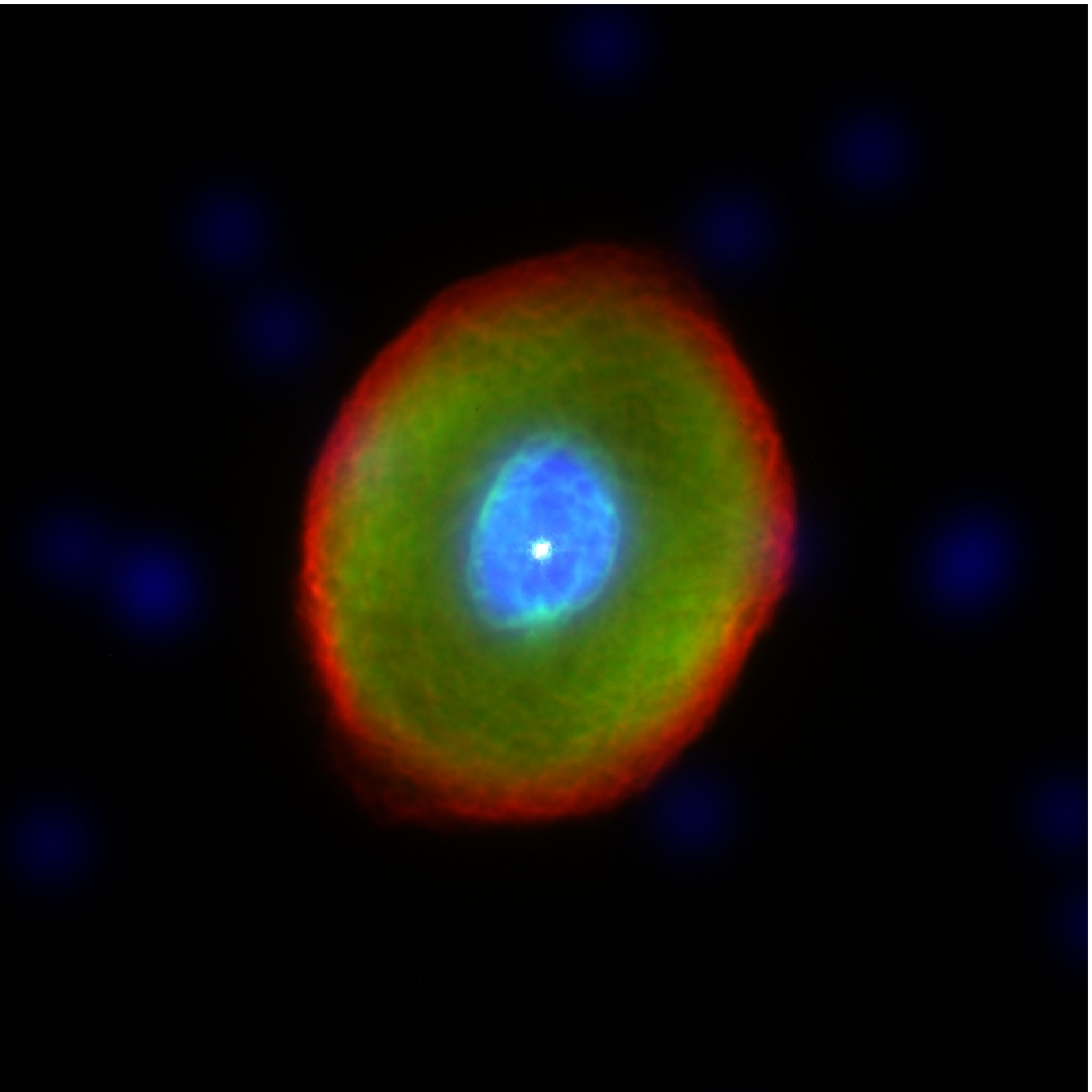} \\
\includegraphics[angle=0.0,width=0.76\columnwidth]{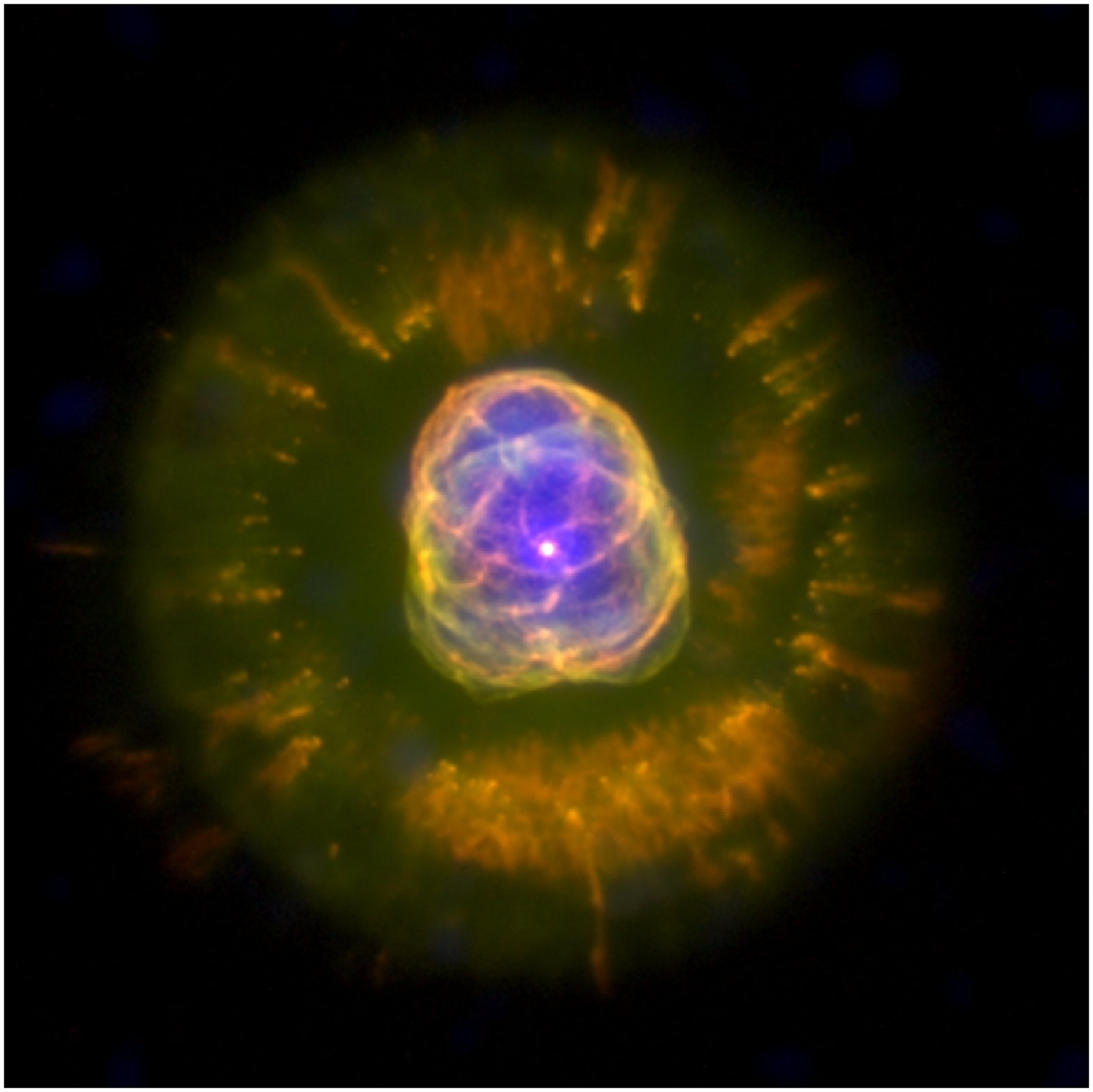} \\
\includegraphics[angle=0.0,width=0.76\columnwidth]{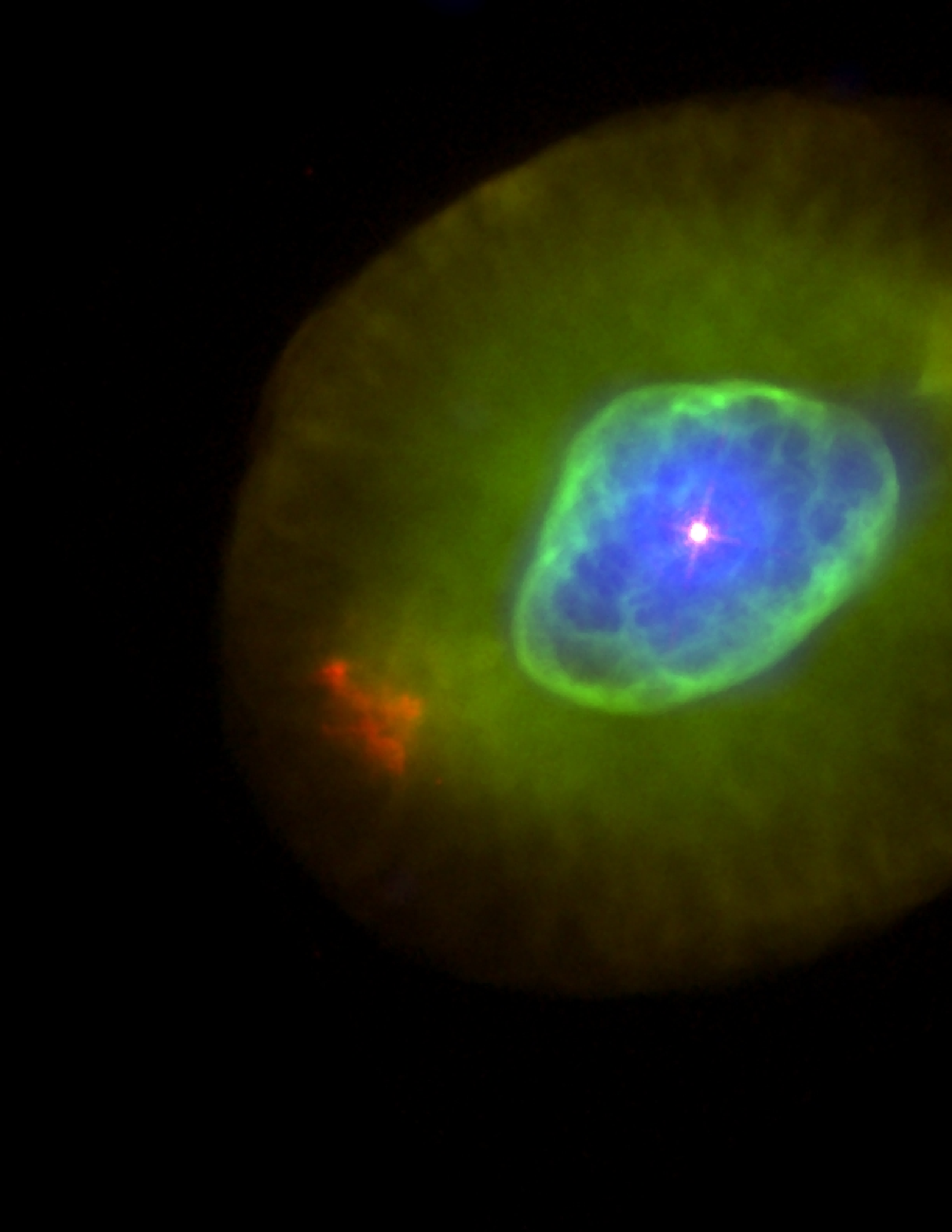}
\caption{
\emph{Chandra} and \emph{HST} composite pictures of IC\,418 ({\it upper}), 
NGC\,2392 ({\it center}), and NGC\,6826 ({\it bottom}).  
X-ray emission is shown in blue, and optical H$\alpha$ and [N~{\sc ii}] 
line emission in green and red, respectively.  
The field of view (FoV) of the pictures is $\approx$27\arcsec\ for IC\,418, 
$\approx$63\arcsec\ for NGC\,2392, and $\approx$41\arcsec\ for NGC\,6826.  
}
\label{color}
\end{figure}

\section{Chandra observations}

The \emph{Chandra X-ray Observatory} (\emph{CXO}) observed IC\,418, 
NGC\,2392, and NGC\,6826 using the array for spectroscopy of the 
Advanced CCD Imaging Spectrometer (ACIS-S).  
All three PNe were imaged on the back-illuminated CCD 
S3 using the VFAINT mode.  
Details of the observations are given in Table~\ref{obs}.  
The \emph{Chandra} observations of NGC\,6826 were split 
into two segments due to scheduling issues.

\begin{figure}[!t]
\centering
\includegraphics[bb=35 168 552 590,width=0.93\columnwidth]{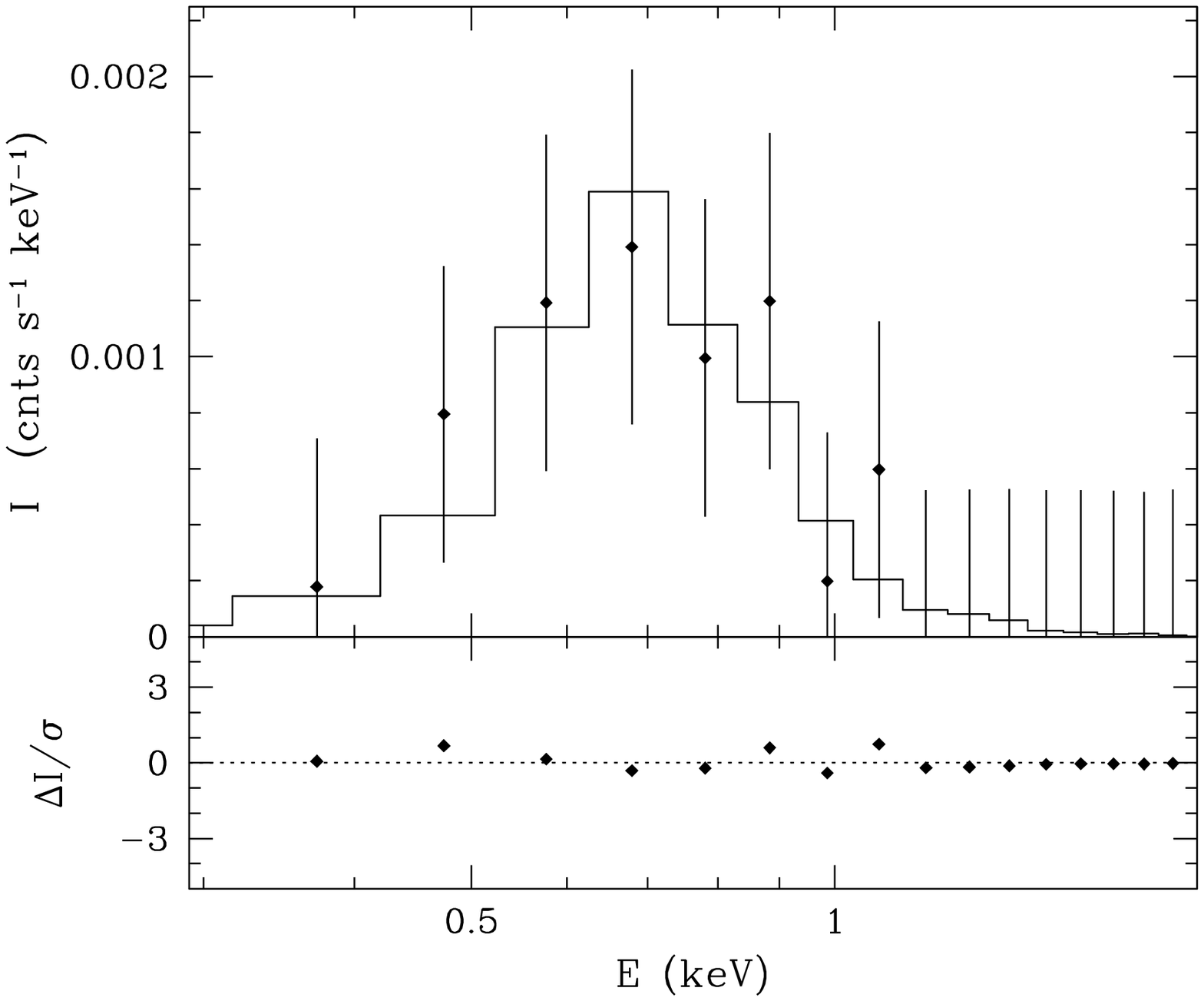}
\\
\includegraphics[bb=35 168 552 590,width=0.93\columnwidth]{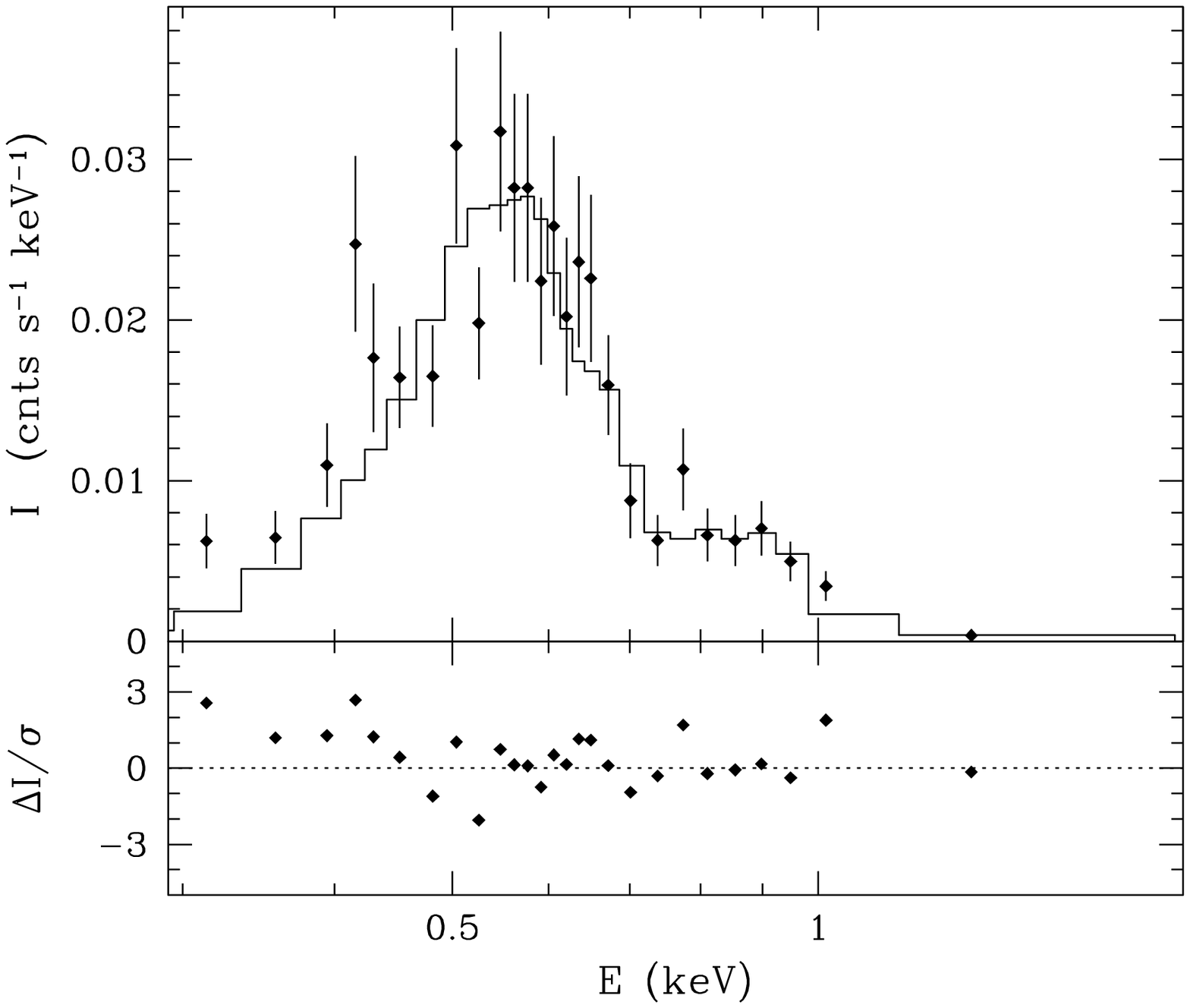}
\\
\includegraphics[bb=35 168 552 590,width=0.93\columnwidth]{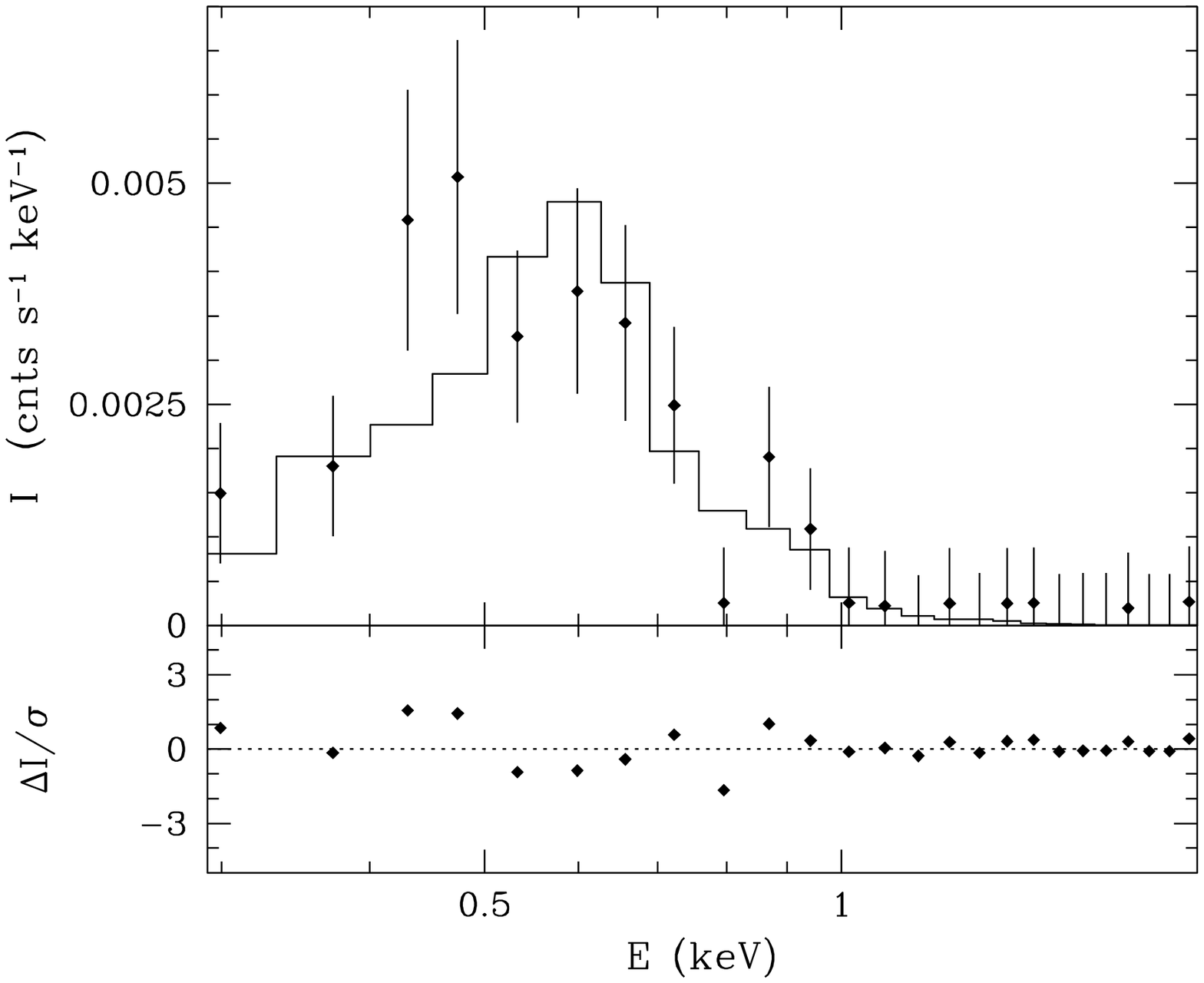}
\caption{
\emph{Chandra} ACIS-S S3 background-subtracted spectra of the diffuse 
X-ray emission of IC\,418 ({\it upper}), NGC\,2392 ({\it center}), and 
NGC\,6826 ({\it bottom}).  
The histograms correspond to models that describe well the observed 
X-ray spectra of IC\,418 and NGC\,6826, while it is the best-fit model 
for NGC\,2392.  
}
\label{spec}
\end{figure}

The background emission level was mostly stable for the observations 
of the three PNe, with NGC\,2392 having a quiescent background level, 
and IC\,418 and NGC\,6826 having abnormally high background levels.  
Besides a short spike of high background that affected the observation of 
IC\,418, no further period has been excised from the original data, 
and therefore the differences between the observation times $t_{\rm obs}$ 
and the useful exposure times $t_{\rm exp}$ in Table~\ref{obs} are mostly 
associated with the removal of dead-time periods.  
All subsequent analyses have been performed using the \emph{Chandra} 
Interactive Analysis of Observations (CIAO) software package version 
4.1.2 and HEASARC XSPEC v12.3.0 routines \citep{A96}.

The \emph{Chandra} ACIS-S X-ray images of IC\,418, NGC\,2392, and 
NGC\,6826 are displayed in Figures~\ref{img} and \ref{color}.  
The raw X-ray images in Figure~\ref{img} have been extracted using the 
natural ACIS-S pixel size of 0\farcs5.  
The smoothed images in Figures~\ref{img} and \ref{color} have been 
processed using the CIAO task \emph{``csmooth''} with a circular 
Gaussian kernel with size up to 4 pixels ($\approx$2\farcs0) and a 
fast-Fourier transform (FFT) convolution method.  
These images reveal the presence of soft diffuse X-ray emission 
in the three PNe.  
A comparison with archival \emph{HST} WFPC2 narrow-band optical 
images selected to emphasize the innermost shells and small-scale 
nebular features (Tab.~\ref{obs}) shows that this diffuse X-ray 
emission is confined within the innermost nebular shell 
(Figures~\ref{img} and \ref{color}).  
We describe below in more detail the spatial and spectral 
properties of this diffuse emission.  
The analysis of X-ray point-sources at the CSPNe of NGC\,2392 and NGC\,6826 
will be reported in an upcoming paper (Guerrero et al., in preparation).

\subsection{IC\,418}

The \emph{Chandra} observations of IC\,418 detect diffuse emission at 
an ACIS-S S3 count rate of 0.62$\pm$0.12 cnt ks$^{-1}$ in the energy 
range 0.3-2.0 keV for a total of 32$\pm$6 counts.  
The X-ray emission is confined within the high excitation, 
innermost 3\farcs5$\times$5\farcs0 shell traced by the 
emission in the [O~{\sc iii}] line (Figs.~\ref{img} and \ref{color}).  
Owing to the small angular size of the X-ray-emitting region and 
small number of counts, the spatial distribution of the X-ray 
emission cannot be analyzed in detail.  
Both the raw image and contours on the top panels of Figure~\ref{img} 
suggest that the diffuse X-ray emission is elongated along the major 
axis of the inner shell.  
The peak of the X-ray emission is located $\sim$0\farcs5 towards 
the northwest of the location of the central star.  
The spatial distribution of the diffuse X-ray emission of IC\,418 relative 
to the optical shell and CSPN is reminiscent of that of BD+30\degr3639 
\citep{Kastner_etal00}.

\begin{deluxetable*}{llclcrcrcl}
\tabletypesize{\scriptsize}
\tablewidth{0pt}
\tablecaption{X-ray Detections of Diffuse Emission from PNe with Bubble Morphology}
\tablehead{
\colhead{Object}         & 
\colhead{Telescope \&}   & 
\colhead{$d$\tablenotemark{a}} & 
\colhead{Hot Bubble}           & 
\colhead{Count Rate}           &
\colhead{$f_{\rm X}$}          & 
\colhead{$S_{\rm X}$\tablenotemark{b}}          & 
\colhead{$L_{\rm X}$\tablenotemark{b}}          & 
\colhead{$kT_{\rm X}$}                 & 
\colhead{Ref.}           \\
\colhead{}                     & 
\colhead{Instrument}           & 
\colhead{}                     & 
\colhead{Radius}               & 
\colhead{}                     & 
\colhead{}                     & 
\colhead{}                     & 
\colhead{}                     & 
\colhead{}                     & 
\colhead{}                     \\ 
\colhead{}                     & 
\colhead{}                     & 
\colhead{[kpc]}                & 
\colhead{[pc]}                 & 
\colhead{[cnt ks$^{-1}$]}          & 
\colhead{[erg\,cm$^{-2}$s$^{-1}$]} & 
\colhead{[erg\,cm$^{-2}$s$^{-1}$arcsec$^{-2}$]} & 
\colhead{[erg\,s$^{-1}$]}          &  
\colhead{[keV]}                    & 
\colhead{}                      
}
\startdata
BD+30$^\circ$3639 & \emph{CXO} ACIS-S & 1.30 &  ~~~0.013 &  244$\pm$4~~~ & 5.7$\times$10$^{-13}$~~ & ~~2.2$\times$10$^{-14}$ & 2.7$\times$10$^{32}$ & 0.233$\pm$0.007    & 1 \\
IC\,418           & \emph{CXO} ACIS-S & 1.20 & ~~~0.012 & 0.62$\pm$0.12 & 2.5$\times$10$^{-15}$~~ & ~~8.4$\times$10$^{-17}$ & 8.4$\times$10$^{29}$ & 0.26               & 2 \\
NGC\,40           & \emph{CXO} ACIS-S & 1.02 & ~~~0.10  &  2.8$\pm$0.9  & 1.3$\times$10$^{-14}$~~ & ~~8.3$\times$10$^{-17}$ & 2.1$\times$10$^{31}$ & 0.09               & 3,4\tablenotemark{c} \\
NGC\,2392         & \emph{CXO} ACIS-S & 1.28 & ~~~0.050 &  9.3$\pm$0.4  & 3.9$\times$10$^{-14}$~~ & ~~1.0$\times$10$^{-16}$ & 1.8$\times$10$^{31}$ & 0.18$\pm$0.04      & 2 \\
NGC\,3242         & \emph{XMM} EPIC   & 1.0~ & ~~~0.041 & 31.3$\pm$1.6~ & 4.2$\times$10$^{-14}$~~ & ~~5.8$\times$10$^{-17}$ & 7.3$\times$10$^{30}$ & 0.202$\pm$0.011    & 5 \\
NGC\,5315         & \emph{CXO} ACIS-S & 2.62 & ~~~0.017 & 12.4$\pm$0.7~ & 1.0$\times$10$^{-13}$~~ & ~~1.5$\times$10$^{-14}$ & 2.9$\times$10$^{32}$ & 0.224$\pm$0.022    & 4 \\
NGC\,6543         & \emph{CXO} ACIS-S & 1.50 & ~~~0.037 & 42.4$\pm$0.9~  & 9.5$\times$10$^{-14}$~~ & ~~5.7$\times$10$^{-16}$ & 6.5$\times$10$^{31}$ & 0.145$\pm$0.010    & 6 \\
NGC\,6826         & \emph{CXO} ACIS-S & 1.30 & ~~~0.028 & 1.97$\pm$0.21 & 9.0$\times$10$^{-15}$~~ & ~~3.4$\times$10$^{-17}$ & 2.0$\times$10$^{30}$ & 0.20               & 2 \\
NGC\,7009         & \emph{XMM} EPIC   & 1.45 & ~~~0.051 & 61.5$\pm$1.7~ & 7.2$\times$10$^{-14}$~~ & ~~1.9$\times$10$^{-16}$ & 4.4$\times$10$^{31}$ & 0.152$\pm$0.015    & 7 \\
NGC\,7027         & \emph{CXO} ACIS-S & 0.89 & ~~~0.012 & 14.0$\pm$0.9~ & 3.1$\times$10$^{-14}$~~ & ~~5.2$\times$10$^{-14}$ & 1.3$\times$10$^{32}$ & 0.26                & 8 
\enddata 
\label{XPN}
\tablenotetext{a}{
Distances adopted from \citet{Mellema04} and \citet{Frew08}.}
\tablenotetext{b}{
Unabsorbed X-ray surface brightness and luminosity in the energy 
range 0.3-2.0 keV. }
\tablenotetext{c}{
\citet{MTZ05} mistakenly reported the absorbed X-ray luminosity of 
NGC\,40 whose value was later corrected by \citet{KN08}.
}
\tablerefs{
(1) \citet{Kastner_etal00}, 
(2) this work, 
(3) \citet{MTZ05}, 
(4) \citet{KN08}, 
(5) \citet{Ruiz_etal11}, 
(6) \citet{Chu_etal01}, 
(7) \citet{Guerrero_etal2005}, 
(8) \citet{Kastner_etal01}. 
}
\end{deluxetable*}

The ACIS-S S3 spectrum of IC\,418 peaks at $\simeq$0.7 keV 
(Figure~\ref{spec}).  
A spectral fit is obviously not possible, but the observed spectrum 
can be reasonably well described by a thin plasma emission model 
with a plasma temperature of 0.26 keV and nebular chemical abundances 
absorbed by a column density of $N_{\rm H}$ $\sim$ 
1$\times$10$^{21}$ cm$^{-2}$ \citep{Pottasch_etal2004}.
In this model, the observed X-ray flux in the 0.3-2.0 keV band is 
2.5$\times$10$^{-15}$ erg~cm$^{-2}$~s$^{-1}$, and the intrinsic 
X-ray luminosity is 8.4$\times$10$^{29}$ erg~s$^{-1}$ for a distance 
of 1.2 kpc \citep{Frew08}.  
These values are compiled in Table~\ref{XPN}, which includes other X-ray 
properties such as the hot bubble radius and unabsorbed X-ray surface 
brightness.

\subsection{NGC\,2392}

The \emph{Chandra} observations of NGC\,2392 resolve the X-ray emission  
detected by previous \emph{XMM-Newton} observations \citep{Guerrero_etal2005} 
into a point-source at the central star and diffuse emission within the 
innermost shell of this nebula (Figures~\ref{img} and \ref{color}).  
The diffuse emission is well confined within the 
15\farcs2$\times$17\farcs6 innermost nebular shell 
of NGC\,2392.  
Its ACIS-S S3 count rate in the energy range 0.3-1.5 keV is 9.3$\pm$0.4 
cnt ks$^{-1}$, and a total of 530$\pm$25 counts are detected.
The diffuse X-ray emission does not show a limb-brightened morphology, 
but it is brighter in the central region and towards a northern 
region inside the central cavity.  
The spatial distribution of this X-ray emission is roughly 
consistent with that expected for an ellipsoidal shell 
filled with hot gas.

The ACIS-S S3 spectrum of the diffuse emission of NGC\,2392 peaks at 0.5-0.6 
keV and shows a plateau of fainter emission between 0.7 and 1.0 keV.  
Very little emission is detected above 1.0 keV.  
The ACIS-S spectrum can be well fit by a thin plasma MEKAL 
emission model.  
The analysis of the \emph{XMM-Newton} data \citep{Guerrero_etal2005} suggested 
that the X-ray-emitting plasma in NGC\,2392 had N/O and Ne/O abundance
ratios greater than the respective nebular values of 0.4 and 0.2 
\citep{Barker1991,HKB2000}.  
The more recent nebular abundance study of NGC\,2392 by 
\citet{PB-SR2008} suggests N/O=0.65 and Ne/O=0.30.  
These nebular abundances, together with the absorption column density 
$N_{\rm H}$=9$\times$10$^{20}$ cm$^{-2}$ converted from the optical 
extinction, provide a reasonably good fit to the observed X-ray 
spectrum (reduced $\chi^2$=1.12) for a plasma temperature of 
$kT$=0.18$\pm$0.04 keV.  
In this model, the observed flux in the 0.2-1.5 keV band is 
3.9$\times$10$^{-14}$ erg~cm$^{-2}$~s$^{-1}$, and the intrinsic 
X-ray luminosity is 1.8$\times$10$^{31}$ erg~s$^{-1}$ for a 
distance of 1.28 kpc \citep{Frew08}.  
Other X-ray properties of NGC\,2392 are listed in Table~\ref{XPN}.

The comparison between the current spectral fit and that performed on the 
\emph{XMM-Newton} data obtained on 2004 April 2 needs to take into 
account that \emph{XMM-Newton} did not resolve the point source at the 
central star from the diffuse emission.  
The physical conditions implied by the spectral fits 
to the diffuse emission detected by \emph{Chandra} 
($N_{\rm H}$=9$\times$10$^{20}$ cm$^{-2}$, $kT$=0.18$\pm$0.04 keV) 
and by \emph{XMM-Newton} 
($N_{\rm H}$=8$\times$10$^{20}$ cm$^{-2}$, $kT$=0.18$^{+0.01}_{-0.03}$ keV) 
are consistent, but the observed diffuse X-ray flux derived from
\emph{Chandra}
($f_{\rm X}$=3.9$\times$10$^{-14}$ erg~cm$^{-2}$~s$^{-1}$) 
is $\sim$35\% lower than that from the \emph{XMM-Newton}
($f_{\rm X}$=6.0$\times$10$^{-14}$ erg~cm$^{-2}$~s$^{-1}$).  
Part of this difference is caused by the contribution of the 
central point source, of which the observed flux is
1.4$\times$10$^{-14}$ erg~cm$^{-2}$~s$^{-1}$ (Guerrero et al., 
in preparation).  
The remaining $\simeq$12\% difference is only slightly larger than the
calibration uncertainties between \emph{Chandra} ACIS and 
\emph{XMM-Newton} EPIC\footnote{
As described by the \emph{XMM-Newton} Calibration Team in 
http://xmm.esac.esa.int/external/xmm\_calibration, 
these differences can amount to 10\%, especially in the soft 
energy range below 1 keV. 
}.

\subsection{NGC\,6826}

The \emph{Chandra} observations of NGC\,6826 detect a point-source 
at its central star (Guerrero et al., in preparation) and diffuse 
emission that, as is the case with IC\,418 and NGC\,2392, is 
confined within the $\sim$12\farcs4$\times$7\farcs4 innermost shell
of the nebula (Figs.~\ref{img} and \ref{color}).  
The diffuse emission is detected at an ACIS-S S3 count rate 
of 1.97$\pm$0.21 cnt ks$^{-1}$ in the energy range 0.3-2.0 keV 
for a total of 96$\pm$10 counts.
The spatial distribution of the diffuse X-ray emission shown by the 
contours on the bottom panels of Figure~\ref{img} hint at a patchy  
distribution which seems to be better described by a centrally filled 
shell than by the limb-brightened morphology that would produce a 
thin shell.

The ACIS-S S3 spectrum of NGC\,6826 is soft, with a plateau of emission 
between 0.35 and 0.7 keV over which a subtle peak at $\sim$0.45 keV can 
be seen.  
A weaker emission peak is detected at $\sim$0.9 keV, and 
no noticeable emission is detected above 1.0 keV.  
The spectral shape is suggestive of emission from an optically thin 
plasma, but, as with IC\,418, no reliable spectral fit is possible due
to the small number of counts.  
Assuming the chemical abundances and absorbing column density of
$N_{\rm H}$ $\sim$ 1$\times$10$^{20}$ cm$^{-2}$ derived for 
NGC\,6826 \citep{SP2008}, the spectrum of its diffuse X-rays can be 
reasonably well described by a plasma emission model for a plasma 
temperature of $kT$ = 0.2 keV and the observed X-ray flux is 
9.0$\times$10$^{-15}$ erg~cm$^{-2}$~s$^{-1}$ in the 0.3-2.0 keV 
energy band.  
The intrinsic X-ray luminosity for this model and energy band is 
2.0$\times$10$^{30}$ erg~s$^{-1}$ for a distance of 1.3 kpc \citep{Frew08}.  
Further X-ray properties of NGC\,6826 are listed in Table~\ref{XPN}.

\section{Discussion}

The background-subtracted spectra of IC\,418, NGC\,2392 and NGC\,6826 have 
been described using optically-thin thermal plasma models with nebular 
abundances and their basic parameters (temperature and X-ray flux and 
luminosity in the energy band 0.3-2.0 keV) determined in previous 
sections.  
The detection of diffuse X-ray emission from these three PNe, together with 
NGC\,6543 and NGC\,7009 \citep{GRU04,ISC02}, testifies that nebular O~{\sc vi} 
emission and/or absorption is an excellent diagnostic for the presence of hot 
bubbles in PNe with sharp shell morphology.

Similarly, the detection of diffuse X-ray emission in a PN can be 
used to forecast the presence of narrow O~{\sc vi} absorptions in 
the spectrum of its CSPN.  
Among the PNe with diffuse X-ray emission 
\citep[this paper;][]{Kastner_etal12}, there are another four with available 
\emph{FUSE} observations of their central stars in the spectral range of the 
O~{\sc vi} lines, namely BD$+$30$\degr$3639, NGC\,40, NGC\,2371-2, and 
NGC\,7662 \citep{GdM13}.  
A careful scrutiny of these spectra reveals narrow O~{\sc vi} absorptions 
of the $\lambda$1031.9 \AA\ line in NGC\,40, NGC\,2371-2, and possibly in 
NGC\,7662, as well as a possible absorption of the $\lambda$1037.6 \AA\ 
line in NGC\,2371-2.  
As for BD$+$30$\degr$3639, the search for narrow O~{\sc vi} absorptions 
in its \emph{FUSE} spectrum is inconclusive because the large number of 
atomic and H$_2$ absorption lines that dominate its stellar continuum at 
this wavelength range.  
To sum up, 8 out of the 9 PNe with diffuse X-ray emission 
also show O~{\sc vi} narrow absorptions in the stellar 
continuum of their CSPNe.

The firm correlation between diffuse X-ray emission and O~{\sc vi} narrow 
absorptions in the stellar continuum of PNe with sharp shell morphology 
provides strong evidence for a conduction layer between the hot interior 
and the cool nebular shell of PNe.  
The physical structure (how the density and temperature vary with radial 
distance) of this conduction layer and the amount of highly ionized species 
present at this interface (O$^{+5}$, N$^{+4}$) depend on the efficiency of 
thermal conduction, although the O~{\sc vi} luminosity seems rather 
insensitive to those effects \citep{Steffen_etal08}.  
We expect our awarded \emph{HST} STIS observations of the N~{\sc v} and
C~{\sc iv} line emission from the interfaces in NGC\,6543 (PI: M.A.\ Guerrero) 
and in the Wolf-Rayet wind-blown bubble S\,308 (PI: Y.-H.\ Chu) would help 
us obtain more information on the physical structure of conduction layers.


\begin{deluxetable*}{lcrrccccrll}
\tabletypesize{\scriptsize}
\tablewidth{0pt}
\tablecaption{
Properties of the stellar wind and diffuse X-ray emission of PNe} 
\tablehead{
\colhead{Object} &
\colhead{Spectral} & 
\colhead{$T_{\rm eff}$} & 
\colhead{$v_\infty$} & 
\colhead{$\log L/L_\odot$\tablenotemark{a}} & 
\colhead{$\dot{M}$\tablenotemark{a}} & 
\colhead{$\log$($L_{\rm X}$/$L_\star$)\tablenotemark{b}} &
\colhead{$\log$($L_{\rm X}$/$L_{\rm wind}$)\tablenotemark{b}} &
\colhead{$T_{\rm shock}$\tablenotemark{c}} &
\colhead{$T_{\rm X}$/$T_{\rm shock}$} & 
\colhead{Ref.} \\
\colhead{} &
\colhead{Type} &
\colhead{} &
\colhead{} &
\colhead{} &
\colhead{} &
\colhead{} &
\colhead{} &
\colhead{} &
\colhead{} &
\colhead{} \\
\colhead{} &
\colhead{} &
\colhead{[kK]} & 
\colhead{[km~s$^{-1}$]} & 
\colhead{} &
\colhead{[$M_\odot$ yr$^{-1}$]} &
\colhead{} &
\colhead{} &
\colhead{[10$^6$ K]} &
\colhead{} & 
\colhead{}
}
\startdata
BD+30$^\circ$3639  & [WC9]   &  47~ &  700~~~ & 3.77 & 1.8$\times$10$^{-6}$ & --4.93 & --3.01 &  14.8~~ & ~~~0.2 & 1 \\  
IC\,418           & Of(H)   &  39~ &  700~~~ & 3.72 & 3.3$\times$10$^{-8}$ & --7.39 & --3.78 &  7.4~~ & ~~~0.4 & 2 \\
NGC\,40           & [WC8]   &  71~ & 1000~~~ & 3.42 & 1.1$\times$10$^{-6}$ & --5.69 & --4.22 & 30.2~~ & ~~~0.035 & 1 \\ 
NGC\,2392         & O6f     &  45~ &  300~~~ & 3.82 & 8.4$\times$10$^{-9}$ & --6.16 & --1.12 &  1.4~~ & ~~~1.5  & 3 \\
NGC\,3242         & O(H)    &  75~ & 2400~~~ & 3.42 & 3.4$\times$10$^{-9}$ & --6.15 & --2.93 & 87.0~~ & ~~~0.03 & 2 \\  
NGC\,5315         & [WO4]   &  76~ & 2400~~~ & 3.74 & 1.6$\times$10$^{-6}$ & --4.87 & --4.00 & 174.0~~ & ~~~0.015 & 1 \\ 
NGC\,6543         & wels    &  60~ & 1450~~~ & 3.54 & 3.3$\times$10$^{-8}$ & --5.32 & --2.53 & 31.8~~ & ~~~0.05 & 2,3 \\  
NGC\,6826         & O3f(H)  &  44~ & 1200~~~ & 3.81 & 4.7$\times$10$^{-8}$ & --7.10 & --4.03 & 22.0~~ & ~~~0.1  & 2 \\
NGC\,7009         & O(H)    &  87~ & 2770~~~ & 3.56 & 1.2$\times$10$^{-9}$ & --5.51 & --1.82 & 115.9~~ & ~~~0.015 & 4,5,6 \\  
NGC\,7027         & $\dots$ & 198~ & $\dots$~~~~ & 3.89 & $\dots$         & --5.37 & $\dots$ & $\dots$~~~ & ~~~~$\dots$ & 7  
\enddata
\tablenotetext{a}{
Stellar luminosities and mass-loss rates adopted from the references in the 
last column and scaled with the distances given in Table~\ref{XPN}.
}
\tablenotetext{b}{
X-ray luminosity in the 0.3-2.0 keV energy band.  
}
\tablenotetext{c}{
The shock temperature of the [WR] stars BD+30$^\circ$3639, NGC\,40, and 
NGC\,5315 is at given wind velocity about twice that of normal CSPNe, 
according to the larger mean molecular weight of their stellar winds.  
}
\tablerefs{
(1) \citet{Marcolino_etal07}, 
(2) \citet{PHM04}, 
(3) \citet{HB11}, 
(4) \citet{Mendez_etal88}, 
(5) \citet{CRP89}, 
(6) \citet{ISC02}, 
(7) \citet{Latter_etal00}.    
}
\label{FUSE_tab}
\end{deluxetable*}

\subsection{Hot Bubbles of Planetary Nebulae}

\emph{Chandra} and \emph{XMM-Newton} have yielded a number of detections 
of diffuse X-ray emission confined within the innermost closed shells of 
PNe, the so-called hot bubbles.  
Sources with available X-ray luminosities and temperatures in the 
literature\footnote{
Recently, \citet{Kastner_etal12} has reported the detection of diffuse 
X-ray emission in the elliptical PNe NGC\,2371-2 and NGC\,7662, but no 
estimates of their X-ray temperatures and luminosities are available.  
Similarly, diffuse X-ray emission has been detected within the innermost 
shell of IC\,4593 (Guerrero et al., in preparation).  } 
are summarized in Table~\ref{XPN} where we have excluded the bipolar PNe 
Mz\,3 and NGC\,7026 \citep{Guerrero_etal04,GRU06,Clark_etal12}.
Table~\ref{XPN} compiles the X-ray properties of the PNe with bubble 
morphology, including their observed X-ray flux ($f_{\rm X}$) and intrinsic 
surface brightness ($S_{\rm X}$) and luminosity in the energy band 0.3-2.0 
keV, scaled to the distance determined by \citet{Frew08}, as well as the 
X-ray temperature and hot bubble radius\footnote{
We note here that the values of the hot bubble radii of the PNe 
in Table~\ref{XPN} that were analyzed by \citet{KN08} clearly 
differ with their values because they used diameters but quoted 
them as radii.  
}.

\begin{figure*}[!t]
\centering
\includegraphics[bb=66 28 565 350,width=1.0\columnwidth]{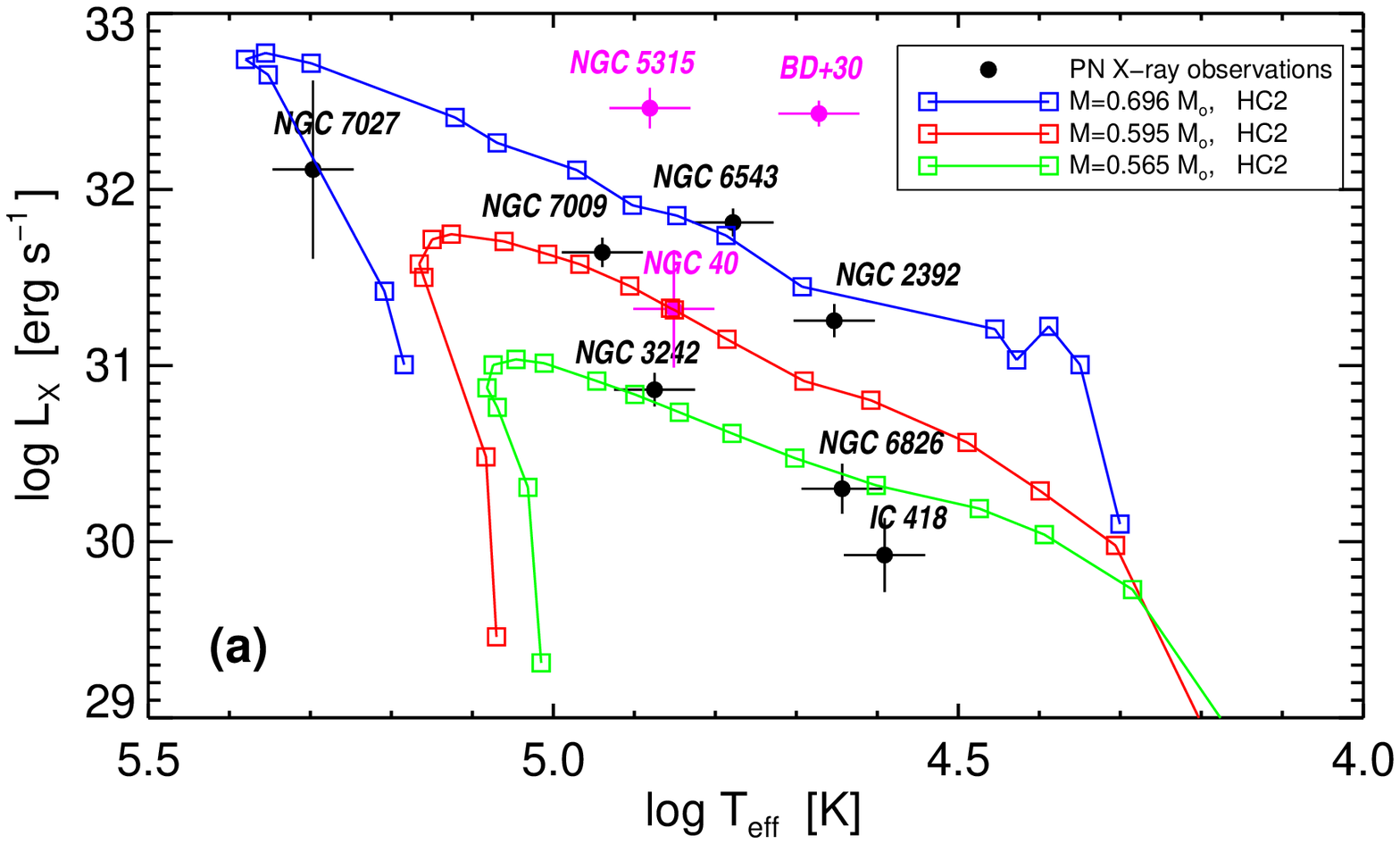}
\includegraphics[bb=56 28 555 350,width=1.0\columnwidth]{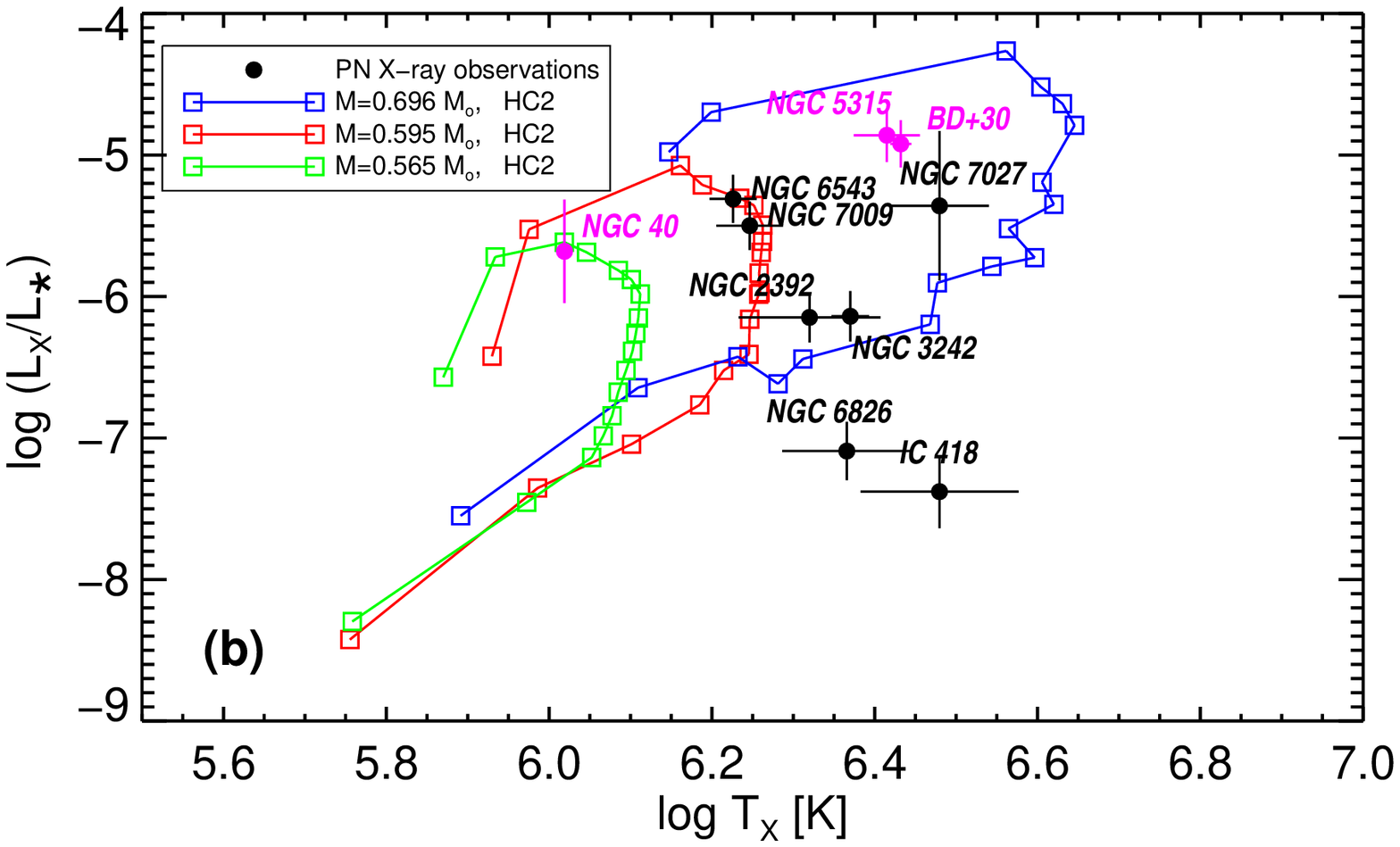}
\\
\includegraphics[bb=66 28 565 350,width=1.0\columnwidth]{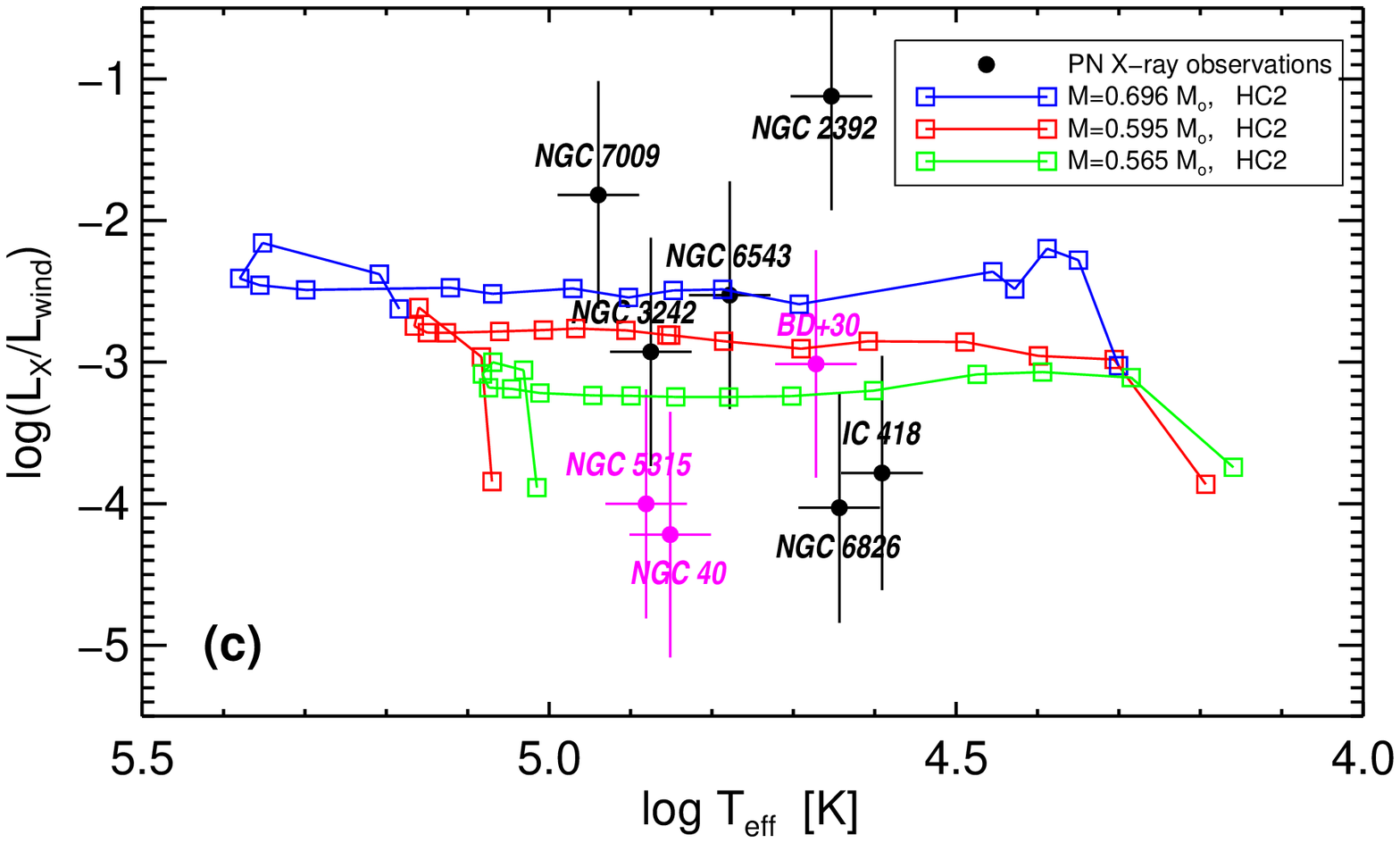}
\includegraphics[bb=56 28 555 350,width=1.0\columnwidth]{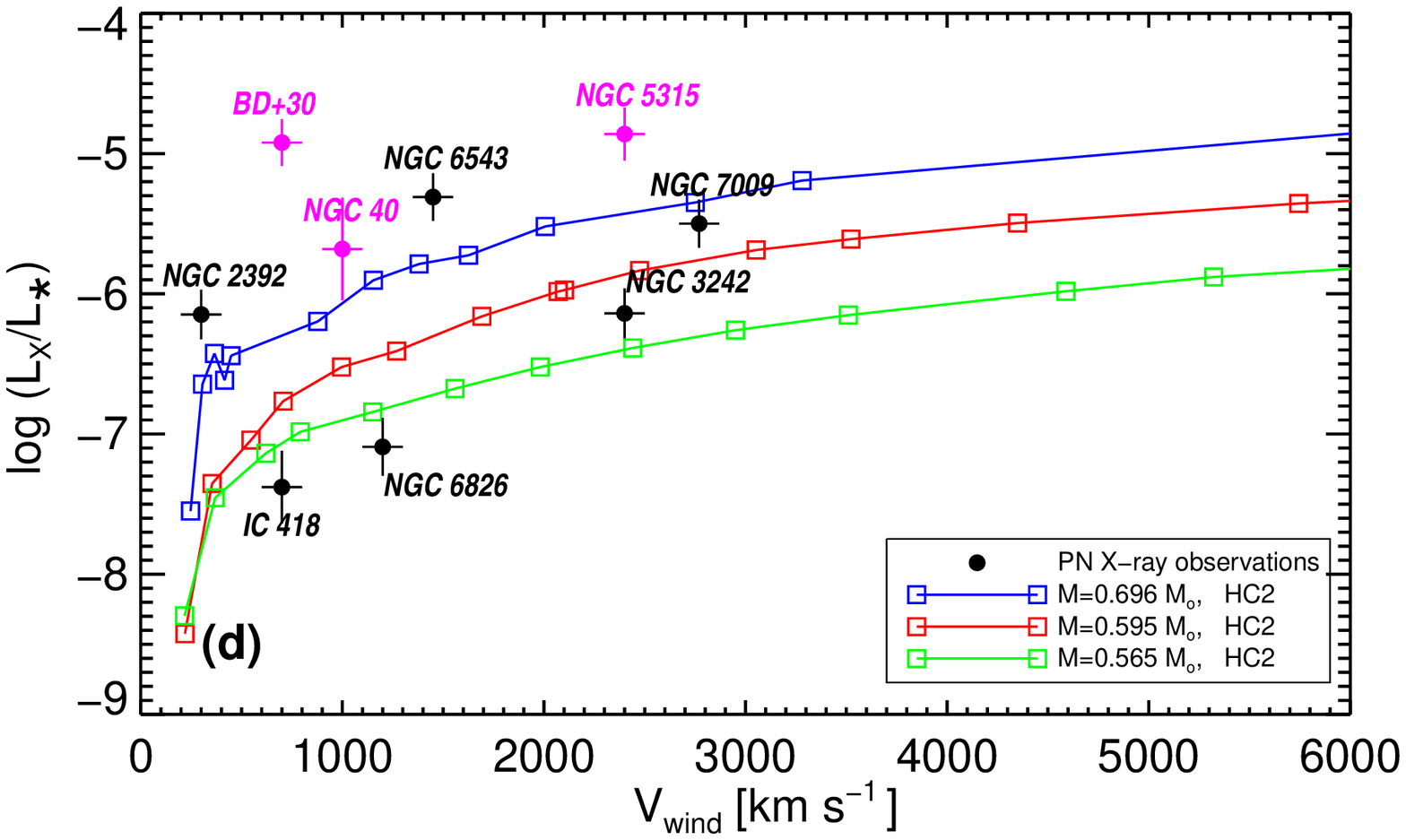}
\\
\includegraphics[bb=66 28 565 350,width=1.0\columnwidth]{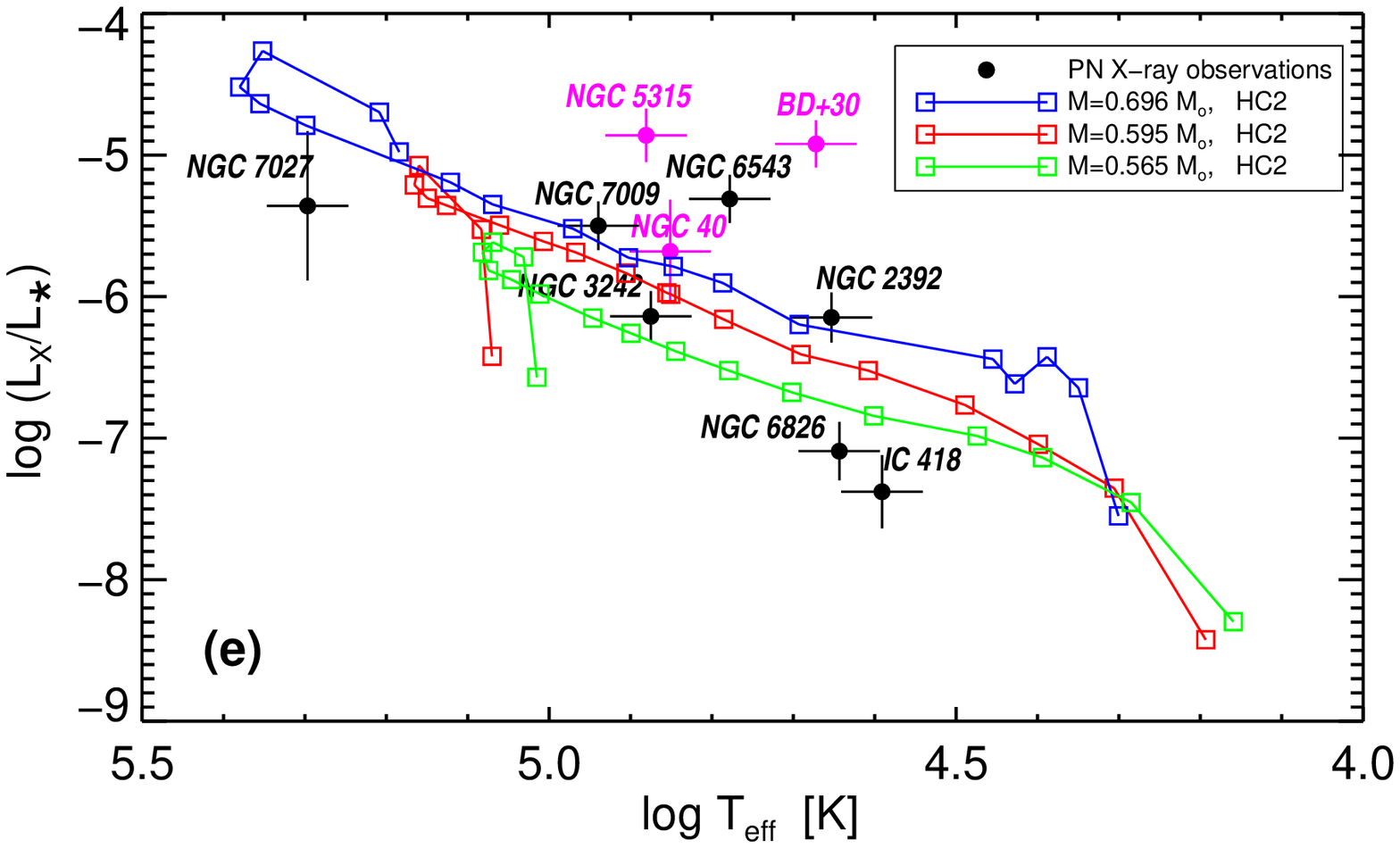}
\includegraphics[bb=56 28 555 350,width=1.0\columnwidth]{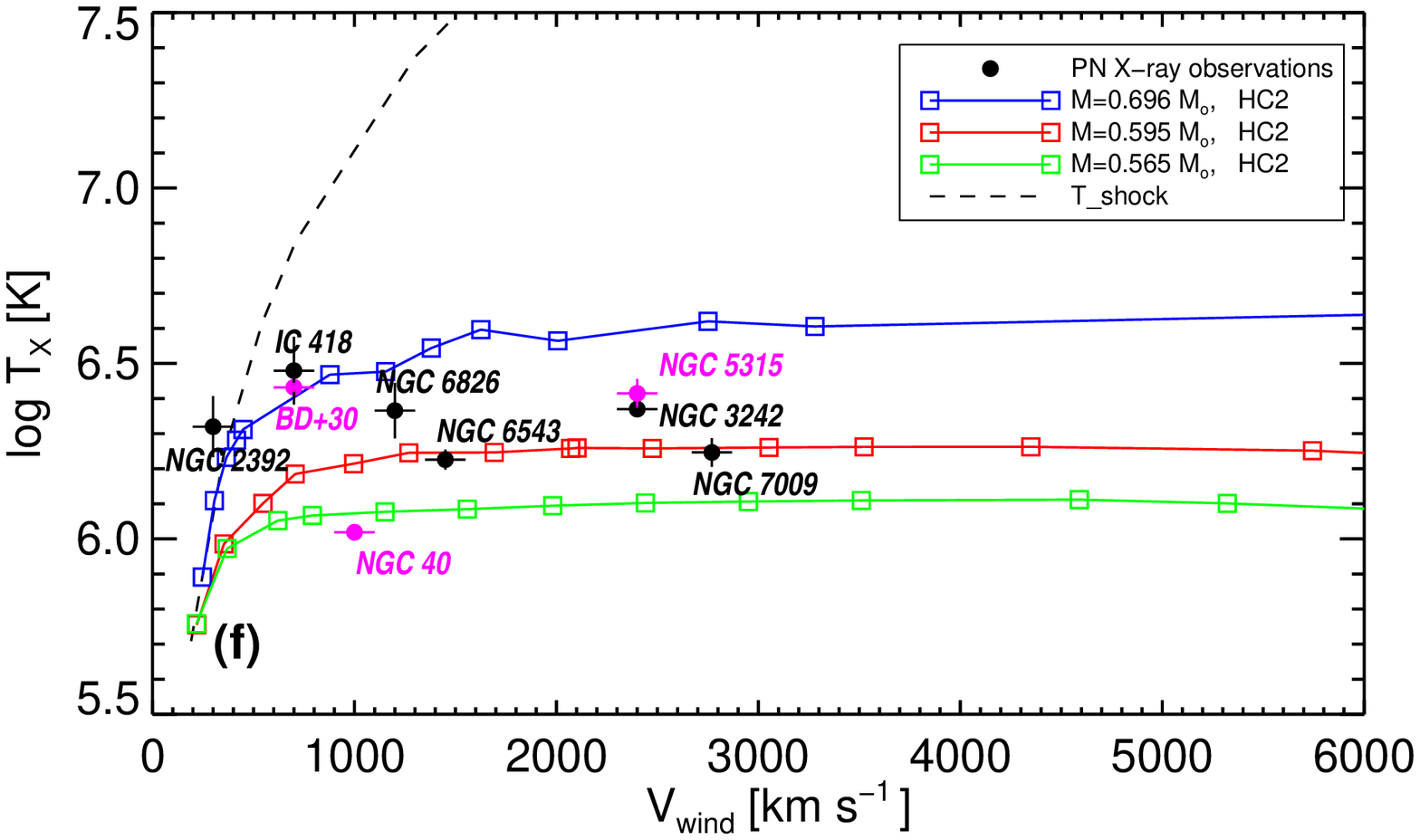}
\caption{
X-ray luminosity ($L_{\rm X}$) of the hot bubble of PNe for the energy 
range 0.3--2.0 keV (6.2--41 \AA) and X-ray temperature ($T_{\rm X}$) as 
functions of $T_{\rm eff}$, the stellar effective temperature (panels a, 
c, and e), X-ray temperature (panel e), and $V_{\rm wind}$, the wind 
terminal velocity (panels d and f).  
In panels b, c,d, and e, $L_{\rm X}$ has been scaled with the stellar 
luminosity ($L_\star$) and wind mechanical luminosity ($L_{\rm wind}$) 
to remove the dependence on distance.  
The tracks correspond to the theoretical models HC2 with heat conduction 
for CSPNe with masses 0.565 $M_\odot$ (green), 0.595 $M_\odot$ (red), and 
0.696 $M_\odot$ (blue) according to the prescriptions of the second method 
used by \citet{Steffen_etal08}, i.e., the heat flux does not exceed the 
saturation limit.  
The dashed line in panel f represents the post-shock temperature of 
an adiabatic shock for the case of H-rich stars, i.e., $T_{\rm shock}$
described by Eq.~1 for $\mu$=0.6.  
Black dots correspond to objects with a H-rich CSPN, 
whereas pink dots refer to sources with a [WR]-type 
CSPN.  
The error-bars in the data points correspond to the 1-$\sigma$ uncertainty in 
our determination of $L_{\rm X}$ and $T_{\rm X}$ (the uncertainty in distance 
is not included in the error-bar of $L_{\rm X}$).  
Typical uncertainties of 0.15 dex for $L_\star$ and 0.8 dex 
for $L_{\rm wind}$ have been convolved with the error-bars of 
$L_{\rm X}$ for the $L_{\rm X}/L_\star$ and $L_{\rm X}/L_{\rm wind}$ 
ratios.  
Similarly, typical uncertainties of 0.05 dex in $T_{\rm eff}$ 
and 100 km~s$^{-1}$ in $V_{\rm wind}$ are presented.  
}
\label{comp}
\end{figure*}

It is interesting to compare the physical properties of these CSPNe and
their winds with the physical parameters of hot gas inferred from the
diffuse X-ray emission.  
The properties of stellar winds of these PNe have been compiled in 
Table~\ref{FUSE_tab} that includes the stellar spectral type (column 
2), effective temperature (column 3), wind terminal velocity (column 
4), stellar luminosity (column 5), and mass-loss rate (column 6) as 
provided by the references listed in the last column of the table.  
An inspection of the spectral type of the CSPNe reveals that this sample 
consists of three H-poor [WR] CSPNe (BD+30$^\circ$3639, NGC\,40, and 
NGC\,5315), and six H-rich CSPNe (IC\,418, NGC\,2392, NGC\,3242, 
NGC\,6543, NGC\,6826, NGC\,7009, and NGC\,7027).  
The wind terminal velocities, stellar luminosities, and mass-loss 
rates have been derived from model atmosphere analyses of UV and 
optical stellar lines, except for NGC\,7027 \citep{Latter_etal00}.  
We note that the stellar luminosities and mass-loss rates 
provided here have been scaled with respect to those given 
in the original references by the distance given in the 
third column of Table~\ref{XPN}.

The X-ray luminosities and temperatures of these nebulae 
are then compared to their stellar ($L_\star$) and wind 
mechanical ($L_{\rm wind} \equiv \frac{1}{2}\;\dot{M}\;v_\infty^2$) 
luminosities in columns 7 and 8 of Table~\ref{FUSE_tab}.  
Finally, the post-shock temperature of the stellar wind ($T_{\rm shock}$), 
estimated assuming an adiabatic shock as 
\begin{equation}
T_{\rm shock} = 
  \frac{3}{16}\,\frac{\mu}{k}\,v^2_\infty \equiv 
  2.3\times10^7\,\mu \left(\frac{v_\infty}{1000~{\rm km\;s}^{-1}}\right)^2 [{\rm K}], 
\end{equation}
where $\mu$ is the mean molecular weight\footnote{
The mean molecular weight for fully ionized H-rich stellar winds can be 
approximated as $\approx$0.6, whereas for fully ionized H-poor stellar 
winds is $\approx$1.2.  
}, is listed in column 9 and 
compared with the plasma temperature derived from X-ray spectra in 
column 10.

These observationally determined properties of hot plasma in
PNe are compared in Figure~\ref{comp} to theoretical predictions based 
on 1D hydrodynamical simulations performed with the code NEBEL 
\citep{Perinotto_etal04,Steffen_etal08,Steffen12}.  
These simulations take into account the energy transfer due to heat 
conduction and the evolution of stellar mass-loss rate and 
terminal wind velocity that determine the mechanical energy 
input into the hot bubble.  
The plots presented in Figure~\ref{comp} are thus an increment and 
refinement of the plots in \citet{Steffen_etal08} for a larger sample
of PNe with consistently revised stellar properties and diffuse X-ray 
parameters.

Before starting a discussion on the location of the PNe in our sample on these 
plots, we would like to note that the models presented in Figure~\ref{comp} 
have been developed exclusively for PNe with H-rich central stars.  
Therefore, the PNe with [WR] central stars BD+30$^\circ$3639, NGC\,40, and 
NGC\,5315 (marked in these plots by ``pink'' symbols) are not expected to 
follow these tracks.  
Indeed, the data points of these sources appear notably far from 
these theoretical tracks, e.g., BD+30\degr3639 and NGC\,5315, 
with young [WR] CSPNe, exhibit the highest values of $L_{\rm X}$ 
(Fig.~\ref{comp}-a) and largest $L_{\rm X}/L_\star$ ratios 
(Fig.~\ref{comp}-b, d, and e).  
Interestingly, the X-ray temperatures predicted by the models for 
H-rich stellar winds do not differ critically from the temperature 
of the hot plasma derived for PNe with [WR] CSPNe (Fig.~\ref{comp}-f), 
despite that they have more massive stellar winds with higher 
mechanical luminosities than their H-rich counterparts.  
Theoretical work is underway to investigate the X-ray properties of 
hot bubbles around [WR] CSPNe \citep{Steffen12}.

\begin{figure}[!t]
\centering
\includegraphics[bb=46 28 560 350,width=1.0\columnwidth]{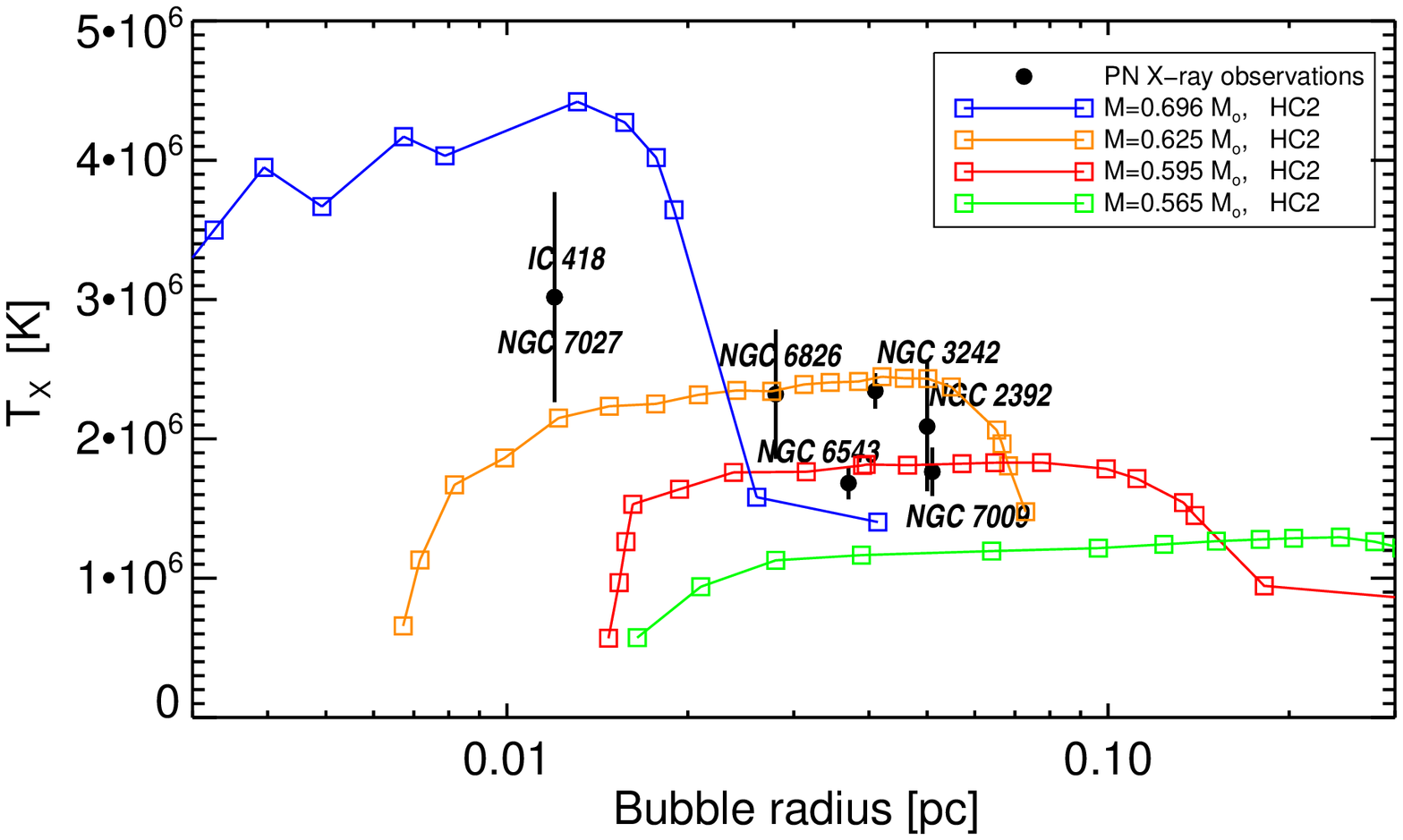}
\includegraphics[bb=46 28 560 350,width=1.0\columnwidth]{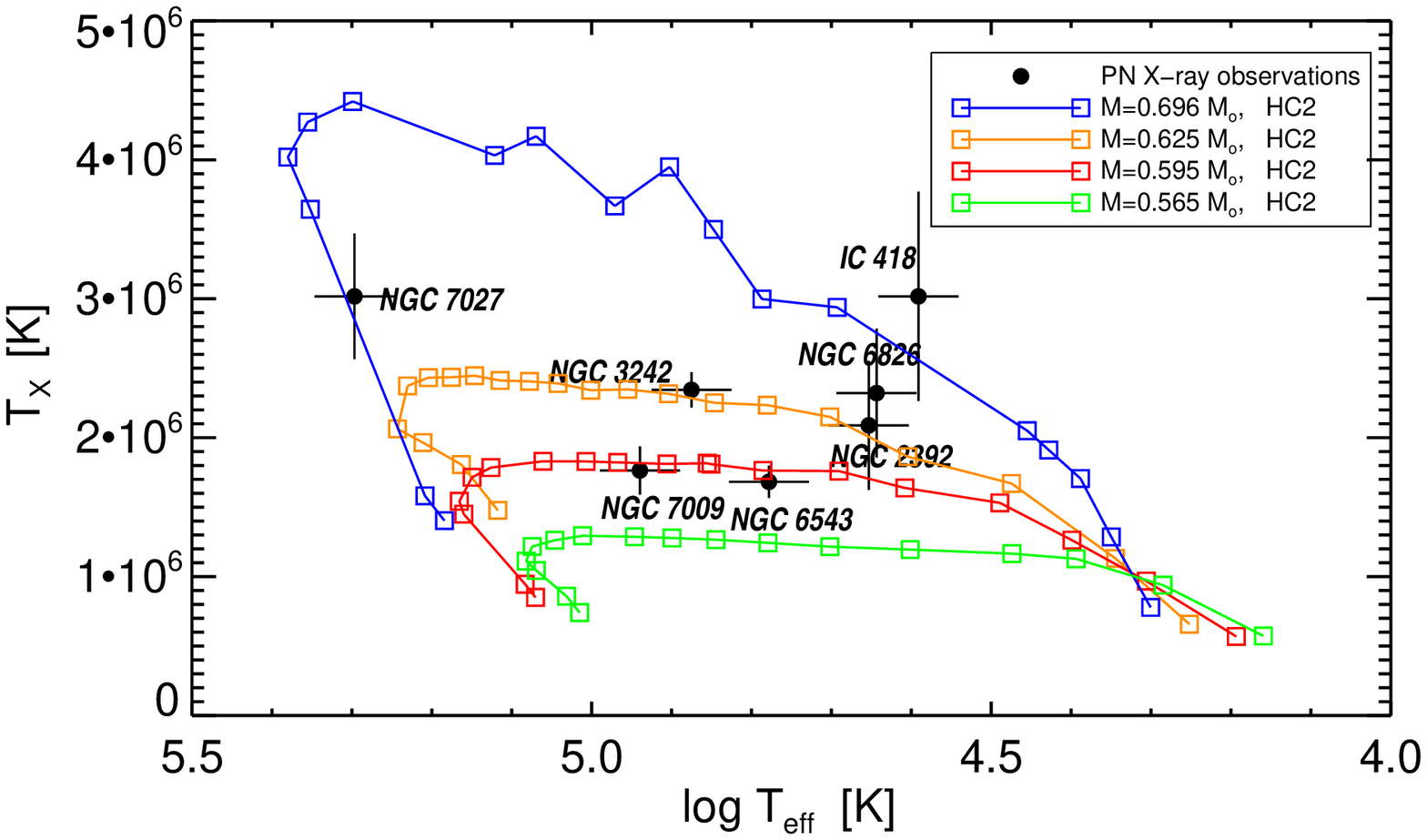}
\includegraphics[bb=46 28 560 350,width=1.0\columnwidth]{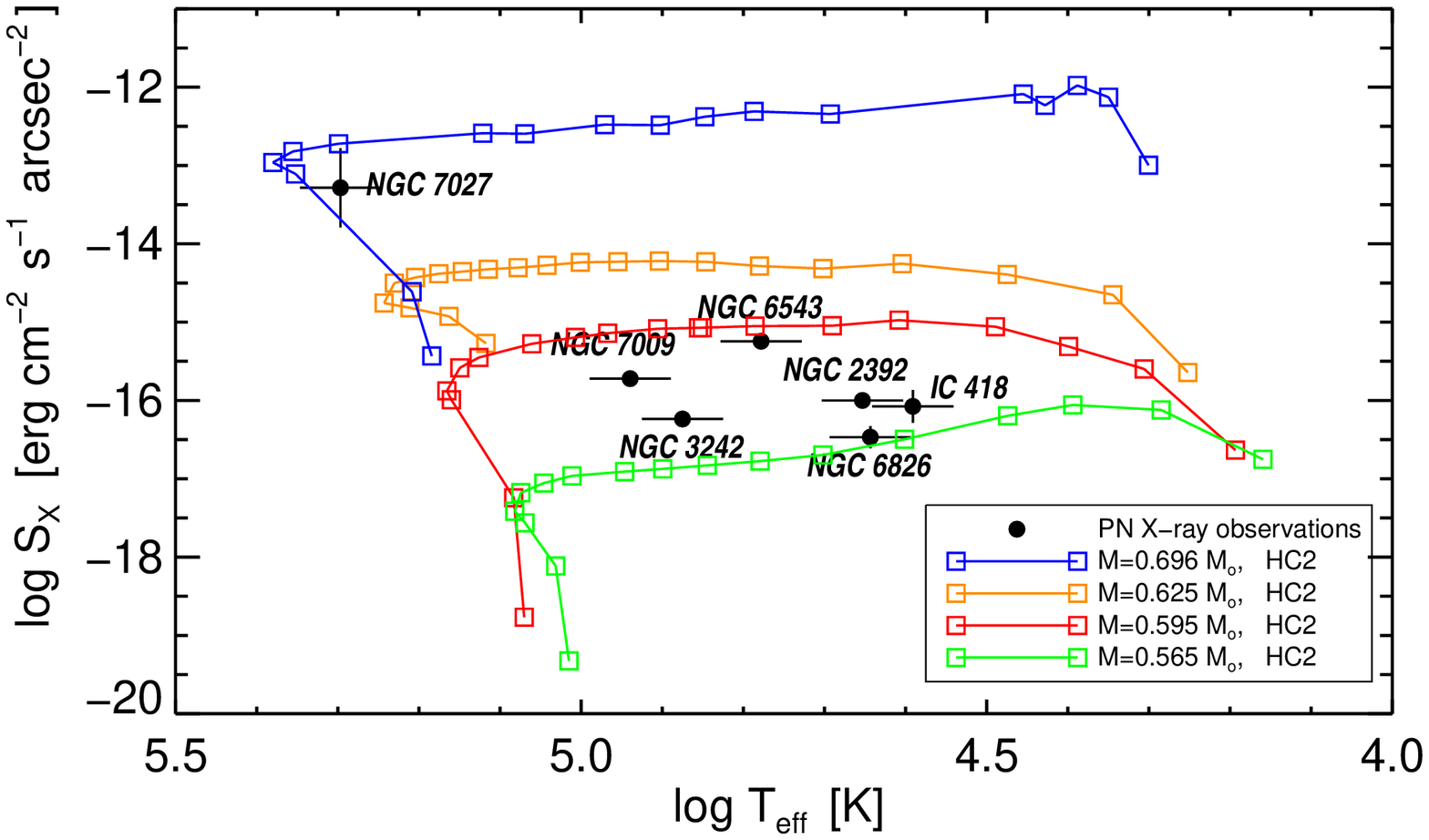}
\caption{
Comparisons of the X-ray plasma temperature with hot bubble radius of PNe
({\it top}) and CSPN effective temperature ({\it middle}), and X-ray surface 
brightness with CSPN effective temperature ({\it bottom}).  
The theoretical tracks of the variations of $T_{\rm X}$ and $S_{\rm X}$ imply 
that the observations comply with the expectations of the theoretical models 
HC2 including heat conduction for central stars with masses 
0.565 $M_\odot$ (green), 0.595 $M_\odot$ (red), 0.625 $M_\odot$ (orange), and 
0.696 $M_\odot$ (blue) adopted from \citet{Steffen_etal08}.  
Error-bars as in Figure~\ref{comp}.
}
\label{hot_bubble}
\end{figure}

Overall, there is a good agreement between the global properties of the 
hot gas in PNe with H-rich stars (``black'' symbols in Figure~\ref{comp}) 
and the model predictions, i.e., most of the data points are bracketed 
within the tracks predicted by the models.  
The most notable differences appear in the comparison between models 
and observations when the stellar wind mechanical luminosity, 
$L_{\rm wind}$, is considered (Fig.~\ref{comp}-c).  
The dispersion of the data points in this figure can stem from the 
large uncertainties in the mass-loss estimates, but also from the 
comparison of the present-time mass loss with X-ray properties that 
depend on the mass-loss rate integrated over time.  
The difficulties for such comparison are well illustrated by NGC\,7009, 
whose low mass-loss rate, as provided by \citet{CRP89}, is increased by 
other authors by factors of $\sim$1.8 \citep{TL02}, $\sim$3.6 
\citep{BKB86}, and $\sim$10.7 \citep{CRP85}.
On the other hand, the increase of X-ray luminosity with time, as 
revealed by the evolution dependent stellar parameters $T_{\rm eff}$ 
and $v_\infty$ shown in panels a, d, and e of Figure~\ref{comp}, is 
consistent with the model predictions.  
Similarly, the little variation of $T_{\rm X}$ with the evolution predicted 
by the models is supported by the data points shown in panels b and f of 
Figure~\ref{comp}.

As for the PNe with new data presented in this paper, we note that 
IC\,418 and NGC\,6826 seem to have low values of $L_{\rm X}$ and 
$L_{\rm X}/L_\star$, and very notably small $L_{\rm X}/L_{\rm wind}$ 
(Fig.~\ref{comp}-c).  
We should emphasize that the X-ray luminosities of these two 
PNe have not been derived from spectral fits, but from a rough 
estimate of their X-ray temperatures and count rates due to 
the small number of detected counts ($\sim$32 cnts and $\sim$96 
cnts, respectively).  
The uncertainty in the X-ray temperature of these sources 
can amount up to 15--25\%, whereas the count rate uncertainty 
is $\sim$20\% for IC\,418 and $\sim$10\% for NGC\,6826.  
Taking into account the uncertainties in count rates, plasma temperatures,
and absorption column densities, the X-ray luminosities of these two 
sources shown in Table~\ref{XPN} can be higher by 50\%.  
Even then, the locations of these two sources in the 
$L_{\rm X}/L_{\rm wind}$ vs.\ $T_{\rm eff}$ (Fig.~\ref{comp}-b) 
and $L_{\rm X}/L_{\rm wind}$ vs.\ $T_{\rm X}$ (Fig.~\ref{comp}-c) 
plots still fall below the theoretical tracks. 
Contrary to IC\,418 and NGC\,6826, the data point of NGC\,2392 is well above 
the theoretical tracks in the $L_{\rm X}/L_{\rm wind}$ vs.\ $T_{\rm X}$ plot 
(Fig.~\ref{comp}-c).  
We will discuss later in section 4.2 the X-ray peculiarities of NGC\,2392.

It has been suggested that the time-scales for the effects of 
stellar wind shock-heated plasma in PNe is short, $\lsim$5,000 yr 
\citep{Ruiz_etal11,Kastner_etal12}.  
To investigate the time evolution of the hot bubbles of PNe, we have plot in 
Figure~\ref{hot_bubble} the X-ray temperature of the PNe in our sample with 
H-rich stellar winds against the radius of their hot bubbles (top panel) and 
the effective temperature of their CSPNe (middle panel).  
The top panel of this figure shows an apparent anti-correlation 
between the X-ray temperature and hot bubble radius, i.e., as 
the hot bubble expands, the X-ray temperature of the plasma seems 
to decrease.  
The comparison of the data points with the theoretical tracks on this 
plot reveals that we may be comparing PNe descending from progenitors 
of different initial masses.  
This is more clearly illustrated in the middle panel plotting 
$T_{\rm X}$ against $T_{\rm eff}$;  PNe with more massive central 
stars produce hotter X-ray-emitting plasma because they reach 
a high wind power in a short time-scale, when the hot bubble 
is still small in size.  
The plot of X-ray surface brightness ($S_{\rm X}$) against the star 
effective temperature, in the bottom panel of Figure~\ref{hot_bubble},
further supports this conclusion.  
The value of $S_{\rm X}$ does not vary much as the star moves along 
the horizontal track of the post-AGB evolution, but it depends strongly 
on the mass of the central star.  
PNe with massive central stars rapidly inject larger amounts of wind 
mechanical energy into small-sized bubbles, resulting not only in higher 
plasma temperatures, but also in higher X-ray surface brightnesses.

Finally, it is interesting to note the persistent absence of limb-brightening 
in the spatial distribution of diffuse X-ray emission from hot bubbles of PNe 
\citep[see also][]{Kastner_etal12}.  
In particular, the diffuse X-ray emission from IC\,418, NGC\,2392, 
and NGC\,6826 does not show clear evidence for limb-brightening 
(Fig.~\ref{img}), suggesting that the X-ray-emitting plasma may fill 
a significant fraction of the nebular shell.  
Whereas models of hot bubbles with heat conduction predict a limb-brightened 
morphology for this diffuse emission, it has been noted that even a small 
amount of interstellar extinction may reduce significantly the center-to-limb 
emission contrast because the soft emission from the cooler plasma close to 
the nebular rim is more easily absorbed than the harder emission from the 
hotter plasma inwards \citep{Steffen_etal08}.  
In this respect, the spatial distribution of diffuse X-ray emission 
from PNe differs notably from that observed in wind-blown bubbles 
around Wolf-Rayet stars, which show distinct limb-brightened morphologies
\citep[e.g., S\,308,][]{Toala_etal12}.

\subsection{NGC\,2392 -- Over-luminous for Its Wind}

An inspection of the location of the PNe whose \emph{Chandra} observations 
are presented in this paper on the different plots shown in Fig.~\ref{comp} 
suggests that IC\,418 and NGC\,6826 are generally a bit under-luminous, 
whereas NGC\,2392 follows the theoretical tracks with the notable exception 
of its high $L_{\rm X}$/$L_{\rm wind}$ in Fig.~\ref{comp}-{\it c}.  
Such peculiar behavior of NGC\,2392 is confirmed by comparing its 
values of $L_{\rm X}$/$L_{\rm wind}$ and $T_{\rm X}$/$T_{\rm shock}$ 
to those of the other PNe in Table~\ref{FUSE_tab}.  
The diffuse X-ray luminosities of these PNe comprise just minute fractions 
of their stellar wind mechanical luminosities; however, the diffuse X-ray 
luminosity of NGC\,2392 constitutes a significant fraction, $\sim$10\%, of 
its wind mechanical luminosity.  
In absolute terms, the X-ray luminosity of NGC\,2392 exceeds that of 
many PNe in Table~\ref{XPN}, which is puzzling as the stellar wind of 
the CSPN of NGC\,2392 has a relatively low mass-loss rate and a meager 
terminal velocity of 300 km~s$^{-1}$.

The plasma temperature of NGC\,2392 is also puzzling.
While the plasma temperatures implied by X-ray spectra 
of all PNe in Table~\ref{XPN} are lower than the post-shock 
temperatures of their stellar winds, the observed plasma 
temperature of NGC\,2392 is {\sl higher} than the post-shock 
temperature expected from its 300 km~s$^{-1}$ wind terminal velocity.
The influence of heat conduction is expected to lower significantly 
the plasma temperature from the post-shock temperature (dashed line 
in Fig.~\ref{comp}-f), which rises with increasing wind velocity,
as it is certainly the case for NGC\,6826. 
However, at low wind speeds, early in the evolution of the hot bubble, the 
influence of heat conduction on plasma temperature is still small, and 
therefore the plasma temperature is not expected to differ substantially 
from the post-shock temperature of an adiabatic shock.  
Yet, the plasma temperature should not exceed the post-shock temperature.

We conclude that the stellar winds of the PNe in our sample 
are able to power their X-ray-emitting hot gas except for 
NGC\,2392.  
A wind terminal velocity higher than 300 km~s$^{-1}$ is needed
for NGC\,2392 to raise its expected for $L_{\rm X}$ and $T_{\rm X}$. 
We note that larger values of the terminal velocity of this wind 
have been reported by \citet{PHM04} and \citet{KUP06}, but 
\citet{HB11} have examined these values and found them not 
compatible with the P-Cygni profiles seen in \emph{FUSE} 
observations.  
Considering that an unusual 200 km~s$^{-1}$ bipolar outflow
has been reported in NGC\,2392 \citep{GIEetal85} and that 
the nebular ionization and expansion velocity are anomalously 
high, it is possible that the CSPN has a hidden companion and 
their binary interactions\footnote{
Such interactions exclude the faint companion reported by \citet{Ciardullo99} 
to be in a possible binary association because at the distance of 2\farcs65 
from the CSPN of NGC\,2392 it would imply an orbital separation of 3,400 AU.  
}
contribute to the energetics of the nebular interior 
\citep[][M\'endez et al., in preparation]{Danehkar_etal12}.  
\citet{Guerrero_etal2005} suggested that a fraction of the diffuse 
X-rays from NGC\,2392 was associated with this bipolar outflow, an 
idea later pursued theoretically by \citet{Akashi_etal08}, but the 
spatial distribution of the diffuse X-ray emission revealed by 
\emph{Chandra} does not support this hypothesis.  
Future investigations of the CSPN of NGC\,2392 are needed to
solve the puzzle of the high X-ray luminosity of its hot interior.

\section{Summary and Conclusions}

We have obtained \emph{Chandra} X-ray observations of three PNe,
IC\,418, NGC\,2392, and NGC\,6826, that display narrow nebular 
O~{\sc vi} absorption lines superposed on the \emph{FUSE} spectra 
of their CSPNe.  
Diffuse X-ray emission is detected in each of these three PNe within 
the sharp innermost optical shell.
These detections, together with those in NGC\,40, NGC\,2371-2, 
NGC\,6543, NGC\,7009, and NGC\,7662, amount to eight known PNe 
with hot ($\sim10^6$~K) interior gas in contact with the cool 
($\sim10^4$~K) nebular shell whose conduction layers have been 
confirmed by the presence of collisionally produced O~{\sc vi}.  
The physical structure of these PNe is thus consistent with that expected 
from bubble models for H-rich stellar winds including heat conduction 
\citep{Steffen_etal08} which predict a conduction layer with a steep 
temperature gradient between the hot interior and the cool nebular shell.

These models also have specific predictions as for the evolution 
in time of the global X-ray properties of bubble models.  
To test these predictions, we have compiled relevant information on the 
X-ray, stellar, and nebular properties of PNe with a bubble morphology.  
The expectations of bubble models including heat conduction compare 
positively with the present X-ray observations of PNe 
with H-rich stellar winds, but those with H-poor [WR] stars present 
notable discrepancies.  
There is an apparent anti-correlation between X-ray temperature and hot 
bubble radius, but we note that this anti-correlation is cause by differences
in stellar masses rather than evolutionary stages:
PNe with massive central stars are expected to produce hotter 
X-ray-emitting plasma inside small-sized hot bubbles of higher 
X-ray surface brightness.  
On the other hand, PNe with less massive central stars have a slower 
evolution during the post-AGB phase and inject less amounts of 
mechanical wind luminosities into larger hot bubbles, resulting in 
cooler X-ray-emitting plasma.

Finally, we note that the stellar wind of PNe with bubble morphology 
and hot gas confined in their interiors seems capable of powering their 
X-ray emission, except for the notable case of NGC\,2392.  
The low speed and small mass-loss rate of its stellar wind result 
in a low wind mechanical luminosity both in absolute and relative 
terms compared to other PNe.  
We suggest that a binary companion may exist and contribute to high 
energy processes that have provided the additional power to its hot 
gas.

\acknowledgments

Support for this work was provided by the National Aeronautics and Space 
Administration through \emph{FUSE} grant NASA NAG~5-13703 and \emph{Chandra} 
Award Numbers SAO~GO7-8019X and SAO~GO1-12029X issued by the \emph{Chandra} 
X-ray Observatory Center, which is operated by the Smithsonian Astrophysical 
Observatory for and on behalf of the National Aeronautics Space Administration 
under contract NASA8-03060.
N.R.\ and M.A.G.\ acknowledge partial support by grants AYA\,2008-01934 and 
AYA2011-29754-C03-02 of the Spanish Ministerio de Educaci\'on y Ciencia 
(currently Ministerio de Econom\'\i a y Competitividad) co-funded by 
FEDER funds.  
We thank Dr.\ Rodolfo Montez for useful discussion on the X-ray 
luminosity of NGC\,40.



{\it Facilities:} 
\facility{\emph{Chandra} (ACIS-S)}, 
\facility{HST (WFPC2)}, 
\facility{FUSE}.

\end{document}